\documentclass{aa}  
\usepackage{graphicx}
\usepackage{txfonts}
\usepackage{subfig}

\usepackage{natbib,twoopt}
\usepackage[colorlinks=true, linkcolor=blue, urlcolor=blue, citecolor=blue, breaklinks=true]{hyperref} 
\bibpunct{(}{)}{;}{a}{}{,} 
\makeatletter
\newcommandtwoopt{\citeads}[3][][]{\href{http://adsabs.harvard.edu/abs/#3}%
{\def\hyper@linkstart##1##2{}%
\let\hyper@linkend\@empty\citealp[#1][#2]{#3}}}
\newcommandtwoopt{\citepads}[3][][]{\href{http://adsabs.harvard.edu/abs/#3}%
{\def\hyper@linkstart##1##2{}%
\let\hyper@linkend\@empty\citep[#1][#2]{#3}}}
\newcommandtwoopt{\citetads}[3][][]{\href{http://adsabs.harvard.edu/abs/#3}%
{\def\hyper@linkstart##1##2{}%
\let\hyper@linkend\@empty\citet[#1][#2]{#3}}}
\newcommandtwoopt{\citeyearads}[3][][]%
{\href{http://adsabs.harvard.edu/abs/#3}
{\def\hyper@linkstart##1##2{}%
\let\hyper@linkend\@empty\citeyear[#1][#2]{#3}}}
\makeatother

\begin{document} 

   \title{ALMA observations of transient heating in a solar active region}

   \subtitle{}

   \author{J. M. da Silva Santos\inst{1} \and J. de la Cruz Rodríguez\inst{1} \and S. M. White\inst{2} \and J. Leenaarts\inst{1} \and \\ G.~J.~M.~Vissers\inst{1} \and V. H. Hansteen\inst{3,4,5,6}
          }

   \institute{Institute for Solar Physics, Department of Astronomy, Stockholm University, AlbaNova University Centre, SE-106 91 Stockholm, Sweden, \email{joao.dasilva@astro.su.se}
         \and 
            Space Vehicles Directorate, Air Force Research Laboratory, Albuquerque, NM, USA
         \and
            Bay Area Environmental Research Institute, NASA Research Park,  Moffett Field, CA 94035-0001, USA
         \and 
            Lockheed Martin Solar \& Astrophysics Lab, Org. A021S, Bldg. 252, 3251 Hanover Street Palo Alto, CA 94304, USA 
         \and 
            Rosseland Centre for Solar Physics and Institute of Theoretical Astrophysics,  University of Oslo, P.O. Box 1029 Blindern, NO-0315 Oslo, Norway
         \and 
            Institute of Theoretical Astrophysics, University of Oslo, P.O. Box 1029 Blindern, NO0315, Oslo, Norway
             }

   \date{}

 
  \abstract
   {} 
   {We aim to investigate the temperature enhancements and formation heights of solar active-region brightenings such as Ellerman bombs (EBs), ultraviolet bursts (UVBs), and flaring active-region fibrils (FAFs) using interferometric observations in the millimeter (mm) continuum provided by the Atacama Large Millimeter/submillimeter Array (ALMA).}
   {We examined 3\,mm signatures of heating events identified in Solar Dynamics Observatory (SDO) observations of an active region and compared the results with synthetic spectra from a 3D radiative magnetohydrodynamic simulation. We estimated the contribution from the corona to the mm brightness using differential emission measure analysis.}
   {We report the null detection of EBs in the 3\,mm continuum at $\sim$1.2\arcsec~spatial resolution, which is evidence that they are sub-canopy events that do not significantly contribute to heating the upper chromosphere. In contrast, we find the active region to be populated with multiple compact, bright, flickering mm-bursts -- reminiscent of UVBs. The high brightness temperatures of up to $\sim$14\,200\,K in some events have a contribution (up to $\sim$7\%) from the corona. We also detect FAF-like events in the 3\,mm continuum. These events show rapid motions of $>$10\,kK plasma launched with high plane-of-sky velocities ($37-340\rm\,km\,s^{-1}$) from bright kernels. The mm FAFs are the brightest class of warm canopy fibrils that connect magnetic regions of opposite polarities.
   The simulation confirms that ALMA should be able to detect the mm counterparts of UVBs and small flares and thus provide a complementary diagnostic for localized heating in the solar chromosphere.}
   {}

   \keywords{ Sun: atmosphere -- Sun: chromosphere --  Sun: corona -- Sun: UV-radiation --  Sun: radio-radiation -- Sun: activity
               }

   \maketitle
%
\section{Introduction}
\label{section:introduction}

Solar active regions are sites of small-scale, transient heating phenomena that are routinely observed in the ultraviolet and visible wavelength range as relatively short-lived, compact brightenings. Among them are Ellerman bombs (EBs), ultraviolet bursts (UVBs), flaring active-region fibrils (FAFs), nano/microflares (NFs/MFs), and other explosive events \citep[see reviews by][and references therein]{Rutten_2013,2018SSRv..214..120Y}. They are commonly regarded as different manifestations of magnetic reconnection. There is mounting observational evidence that they preferentially occur in regions of magnetic flux emergence and where there is interaction between mixed magnetic polarities in a variety of contexts \citep[e.g.,][]{1987ApJ...323..380P,1994AdSpR..14...13D,1997Natur.386..811I,Chae_1998,2011ApJ...736...71W,2013ApJ...774...32V,2016ApJ...824...96T, 2018A&A...615L...9C}. 

Numerical simulations have shown that the magnetic topology that determines the height of reconnection along with the line of sight to the reconnection site play a role in the visibility of these small-scale heating events across the wavelength spectrum \citep{2019A&A...626A..33H,2019A&A...628A...8P,2020A&A...633A..58O,2020ApJ...891...52S}. For example, strong emission in the wings of H$\alpha$ would imply reconnection in the lower atmosphere, whereas bright \ion{Mg}{II} h and k, \ion{Si}{IV} 1393.7, 1402.8\,\AA~or even extreme-ultraviolet (EUV) emission  would originate in higher layers.
Other key signatures such as recurrence, intermittency, and large flow velocities may be caused by the release of fast-moving sequences of magnetic islands (or plasmoids) and jets \citep[e.g.,][]{1995ApJ...450..422K,2001EP&S...53..473S,2009PhPl...16k2102B,2015ApJ...813...86I,2015ApJ...799...79N,2017ApJ...851L...6R,2019A&A...628A...8P}, while turbulence has also been shown to potentially play a role in triggering fast reconnection in MFs \citep[e.g.,][]{2020ApJ...890L...2C}.

The Atacama Large Millimeter/submillimeter Array \citep[ALMA,][]{2009IEEEP..97.1463W} offers a different view into the properties of small-scale heating events by producing high-cadence interferometric maps of the millimeter (mm) continuum at unprecedented spatial resolution at these long wavelengths. The bulk of the solar emission in the mm, and in particular in ALMA Band 3 (3\,mm or 100\,GHz), comes from the chromosphere and from a range of heights from $\sim$1200 -2000\,km above the average height at which the optical depth of the 500\,nm continuum ($\tau_{500}$) is unity in quiet conditions \citep[e.g.,][]{1981ApJS...45..635V,2015A&A...575A..15L}. The main emission mechanism is thermal bremsstrahlung but nonthermal synchroton may also be detected in large flares \citep[see review by][]{Wedemeyer16}. Because the source function of the mm continuum is given by the Planck function, which is nearly linear in temperature in the Rayleigh-Jeans limit, one would expect ALMA to detect the counterpart of UVBs, FAFs, and similar active region phenomena whenever they cause a significant temperature increase at or above the opaque chromospheric canopy \citep{2017A&A...598A..89R}. Additionally, ALMA observations do not suffer from blurring from scattering that affects the more widely used chromospheric diagnostics such as H$\alpha$ and \ion{Mg}{II} h and k \citep[see review by][]{2017SSRv..210..109D} and can be used to better constrain the temperature stratification of the atmosphere \citep{2018A&A...620A.124D}. 

However, recent studies that used some of the first solar ALMA observations brought up the need to better understand the formation heights of the ALMA bands in active regions \citep{2017ApJ...850...35L,2020A&A...634A..56D,2020arXiv200512717C} as they likely differ from the quiet Sun (QS) where the emission is dominated by acoustic shocks \citep[e.g.,][]{2006A&A...456..697W,2007A&A...471..977W,2020A&A...634A..86P}. Moreover, the contribution functions of the mm continuum may also depend on nonequilibrium ionization/recombination effects in the chromosphere \citep[e.g.,][]{2002ApJ...572..626C,2006ASPC..354..306L,2017A&A...598A..89R,2020ApJ...891L...8M}.
In theory, a contribution from the overlying corona to the mm brightness is also expected but deemed to be small, at least in the QS \citep[e.g.,][]{1992SoPh..141..347W}.

\citet{2017ApJ...841L...5S} reported on the first detection of a localized heating event featuring a plasmoid ejection during the ALMA science verification campaign.
In this paper, we address, in a more systematic way, the visibility of different small-scale heating events such as EBs, NFs and FAFs in the mm continuum using recent ALMA observations taken at higher spatial resolution, and we compare them to a snapshot of a 3D radiative magnetohydrodynamic (r-MHD) simulation of magnetic flux emergence.
With regard to  the interpretation of the observation, we discuss the importance of understanding the contribution functions of the ALMA bands, in particular the contribution from the transition region (TR) and corona in high density condition.

\section{Observations}
\label{section:observations}
\subsection{Data reduction and calibration}

\begin{figure*}
\centering
    \includegraphics[width=\linewidth]{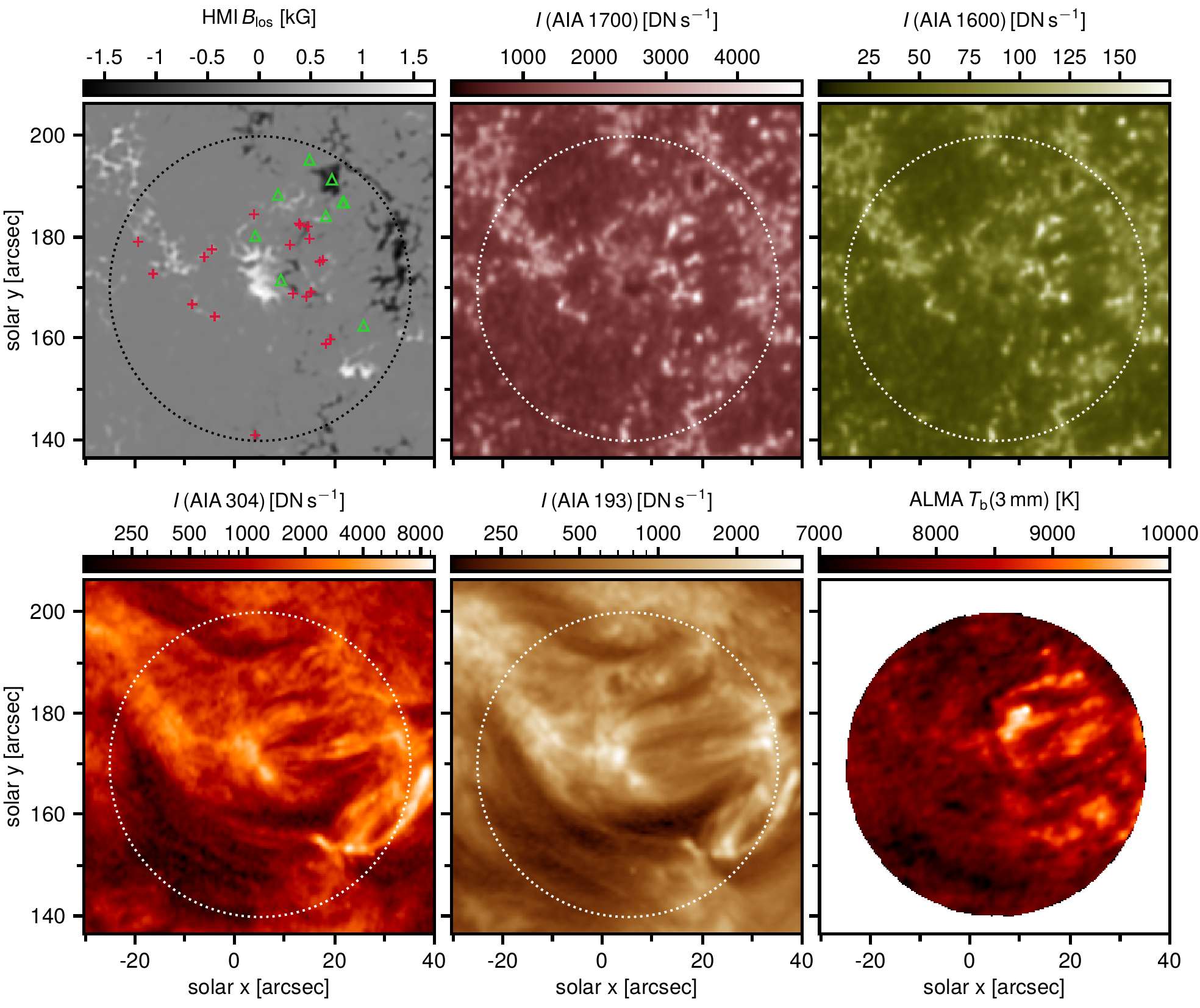}
    \caption{Overview of target as seen by SDO and ALMA on 13 April 2019.
    Clockwise from top left: HMI magnetogram, spectral radiance in AIA 1600\,\AA~and AIA 1700\,\AA, ALMA 3\,mm brightness temperature (capped at 10\,kK for display), and spectral radiance in AIA 193\,\AA~and AIA 304\,\AA. The dotted circle displays the ALMA field of view. The $x$ and $y$ axes are the helioprojective coordinates at approximately 18:20\,UTC. The crosses indicate the location of EBs (see Sect.\,\ref{section:EBsandMFs}), and the triangles correspond to NFs (see Sect.\,\ref{section:DEM}) that we detected in the entire time series.}
    \label{Fig:1}
\end{figure*}

On 13 April 2019, ALMA was pointed at a group of pores in the periphery of a large sunspot in NOAA\,12738 near the disk center ($\mu=0.98$, where $\mu$ is the cosine of the heliocentric angle).
The ALMA Band 3 data were acquired in two execution blocks, each consisting of three 10-minute scans separated by 140-second calibration intervals. Each scan contains 300 2.02-second integrations. The two periods covered by the execution blocks were 18:19:52-18:54:55 and 19:15:32-19:50:31. The array contained 50 antennas (41 $\times$ 12m and 9 $\times$ 7m), of which eight (6 $\times$ 12m, 2 $\times$ 7m) were flagged throughout for various reasons, leaving 42 antennas contributing to the imaging most of the time (other antennas were also occasionally flagged during individual scans).

The data in each execution block were mapped and self-calibrated in phase with  progressively smaller intervals to remove atmospheric effects. The field of view of the 7m antennas is about 100\arcsec\ at 100 GHz, while that of the 12m antennas is 58\arcsec. We made 512$\times$512 pixel maps with a cell size of 0.3\arcsec, but quantitative analysis is restricted to the inner 60\arcsec\ regions of the images. The four sidebands (94, 96, 104, 106 GHz) were included together in the mapping in order to improve instantaneous $u,v$ coverage. The final maps used here were primary-beam-corrected and restored with a Gaussian beam 1.2\arcsec\ in width. The maps are converted from flux to brightness temperature. To put them on an absolute temperature scale, we compared the average temperature within the 60\arcsec\ region of interest in the interferometer maps with the corresponding temperature at that location in a calibrated single-dish image (resolution 60\arcsec). This resulted in the addition of 7700\,K to the interferometer maps. The typical noise level in the difference map made from consecutive maps is 20\,K. The use of phase self-calibration leads to some uncertainty in the absolute positions of the images and can result in the appearance of small artificial motions in a time sequence of images: in the movie of these data this effect is visible, but is too small to affect our analysis.

We also use ultraviolet (UV) images taken with the Atmospheric Imaging Assembly \citep[AIA,][]{2012SoPh..275...17L} and photospheric magnetograms obtained with the Helioseismic and Magnetic Imager \citep[HMI,][]{2012SoPh..275..207S} aboard the Solar Dynamics Observatory
\citep[SDO,][]{2012SoPh..275....3P}.
As part of the routine AIA data reduction, cosmic rays are removed from the EUV images in a so-called de-spiking procedure. While this is desirable for most purposes, it also causes many small-scale, bright events to be erroneously removed.
We therefore "respiked" the level-1 AIA data products using the IDL code \texttt{aia\_respike} in SolarSoftWare \citep[SSW, ][]{1998SoPh..182..497F}. The regions of interest were maintained and the image artifacts caused by cosmic rays were, for the most part, replaced by interpolated values using a Gaussian kernel.
The AIA and HMI data were further processed with routines provided in the \texttt{SunPy} package \citep{2015CS&D....8a4009S}, namely for alignment, scaling, derotation, and resampling. The HMI continuum images and magnetograms were deconvolved and upsampled to 0.3\arcsec per pixel using the \texttt{Enhance} deep learning code\footnote{\url{https://github.com/cdiazbas/enhance}} by \citet{2018A&A...614A...5D} to improve the visibility of small-scale flux emergence in the regions of interest. The AIA EUV data were corrected for instrumental degradation \citep{2014SoPh..289.2377B} and stray light contamination using the semi-empirical point-spread-functions of \citet{2013ApJ...765..144P}. 
Finally, the AIA images were rebinned 0.3\arcsec~to match the deconvolved HMI and ALMA pixel scale.
The cadence of the AIA observations is 12\,s and 24\,s for the EUV and UV passbands, respectively, while that of the HMI magnetograms is 45\,s.

Given the discrepancy between ALMA and SDO coordinates\footnote{Recent work has shown that an inappropriate shift to account for gravitational deflection has been applied to ALMA observations of the Sun during recent Cycles (N. Phillips and R. Marston, ALMA ICT ticket 16261).
This results in a position error that can be as large as 55\arcsec~within 100\arcsec~of the disk center. For this observation, centered 170\arcsec~from the apparent disk center, the offset could be to the order of 10\arcsec.}, the co-alignment was refined by taking advantage of the good correspondence between the mm and EUV bright structures, in particular in the 304\,\AA~passband that, overall, resembles Band\,3 the most in the active region. A particular event at 18:36\,UTC (described in Sect.\,\ref{section:faf}) was especially useful since it features two compact kernels and one bright arch fibril allowing the offset and angle between the different diagnostics to be better constrained. The accuracy was visually verified by the spatial coincidence of other events at different times. However, we do not expect the mm and EUV brightenings to exactly align at all times.
We note that the maximum spatial resolution of these ALMA observations is $\sim$1.2\arcsec and thus similar to AIA.
\subsection{Overview}
\label{section:overview}

\begin{figure*}[t]
    \centering
    \sidecaption
    \includegraphics[width=120mm]{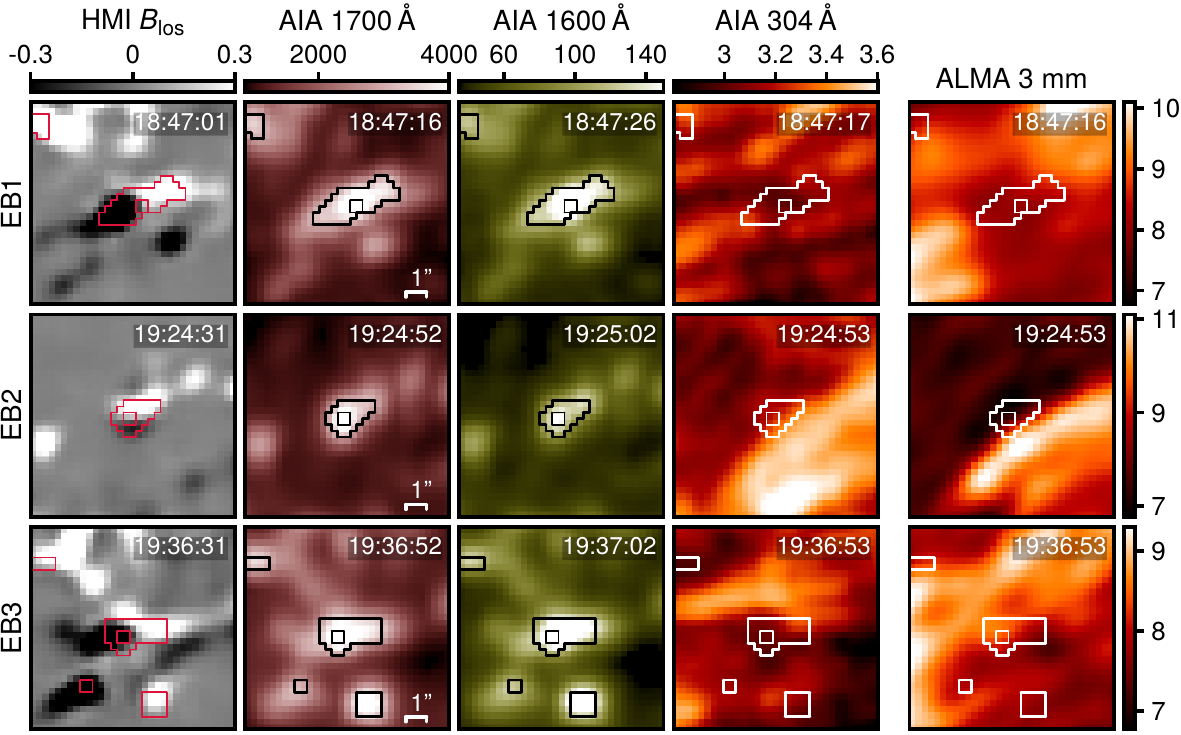} 
    \caption{ Examples of EB candidates in SDO and ALMA. From left to right: HMI magnetogram, spectral radiance in AIA\,1700\,\AA, AIA\,1600\,\AA, and AIA\,304\,\AA, and ALMA 3\,mm brightness temperature.  All panels show a 10\arcsec$\times$10\arcsec~FOV centered on EBs. The contours correspond to $5\sigma$ (thick) and $9\sigma$ (thin) \texttt{EBDETECT} thresholds (Sect.\,\ref{section:EBsandMFs}). The image scale is indicated in the panels in the second column. The range in the HMI magnetograms is clipped at $\pm0.3$\,kG and the AIA images are capped at 4500\,$\rm DN\,s^{-1}$ (1700\,\AA), 150\,$\rm DN\,s^{-1}$ (1600\,\AA)~and 4000\,$\rm DN\,s^{-1}$ (304\,\AA). The 304\,\AA~images are displayed in logarithmic scale. The ALMA color bars are in units of kilokelvin. Additional examples are shown in the supplementary Fig.\,\ref{fig:ebpanelextra}. }
    \label{fig:ebpanel}
\end{figure*}
\begin{figure}
    \centering
    \includegraphics[width=\linewidth]{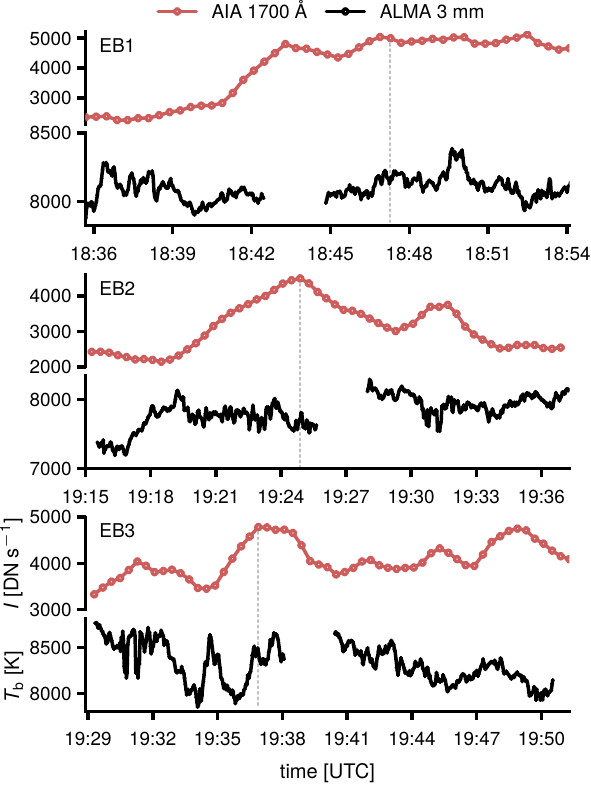}
    \caption{ Light curves of selected EB candidates. From top to bottom: AIA 1700\,\AA~and ALMA 3\,mm light curves at the center of EB1, EB2, and EB3. The vertical dotted lines mark the time stamps displayed in Fig.\,\ref{fig:ebpanel}.}
    \label{fig:eblc}
\end{figure}

Figure\,\ref{Fig:1} shows an overview of the target at the start of the ALMA Band\,3 observations. The right-hand side of the ALMA field of view (FOV) captured an area where significant flux emergence was visible at the time in the vicinity of a group of pores and plage region. The sunspot is located northwest outside of the FOV. The area includes bright, thus warm, compact features and long fibrilar structures that appear to be rooted in the photospheric magnetic elements and connect regions of opposite polarity, similar to an arch-filament system. The latter seem to be a larger-scale analogue of "magnetic loops" reported by \citet{2020A&A...635A..71W}, who show that the 3\,mm continuum brightness correlates with the EUV intensities in these structures.
Other ALMA observations have shown bright fibril-like structures emanating from magnetic concentrations \citep[][]{2019ApJ...881...99M,2020A&A...634A..56D,2020arXiv200512717C}. These features contrast with the weakly magnetized areas in the left-hand side of the FOV that shows lower amplitude variations of $T_{\rm b}$.

The highest $T_{\rm b}$ at the time shown in Fig.\,\ref{Fig:1} is $\approx$10300\,K, but values up to $\approx$14\,300\,K are detected at later times (Sect.\,\ref{section:DEM}). The lowest $T_{\rm b}$ in the entire time series is approximately $\sim$6780\,K in the lower left quadrant of the FOV, and we do not detect "chromospheric holes" as recently found in the QS \citep{2019ApJ...877L..26L} and near active-region plage \citep{2020A&A...634A..56D}. We also note the weak correlation (correlation coefficient $r=0.34$) between AIA 1600\,\AA~and 3\,mm brightness contrary to what has been reported in QS \citep{2018A&A...619L...6N}. 

We identified different types of localized, transient brightenings in the AIA images throughout the duration of the ALMA campaign and we searched for their counterparts in the Band\,3 data. We plot the locations of EBs and NFs on the HMI magnetogram in Fig.\,\ref{Fig:1}. They predominantly occur between the two pores and plage. The EB detections are presented in Sect.\,\ref{section:EBsandMFs} and the NFs are discussed in Sect.\,\ref{section:DEM}. We also observe FAF-like events in the EUV and mm (Sect.\,\ref{section:faf}).
We note that the 3\,mm brightenings described in this paper correspond to much higher amplitude $T_{\rm b}$ variations than the QS transients observed with Band\,3 \citep{2020A&A...638A..62N} and are likely of different origin.

\section{Results}
\label{section:results}

\subsection{Ellerman bombs as sub-canopy events}
\label{section:EBsandMFs}

\begin{figure*}
    \centering
    \sidecaption
    \includegraphics[width=120mm]{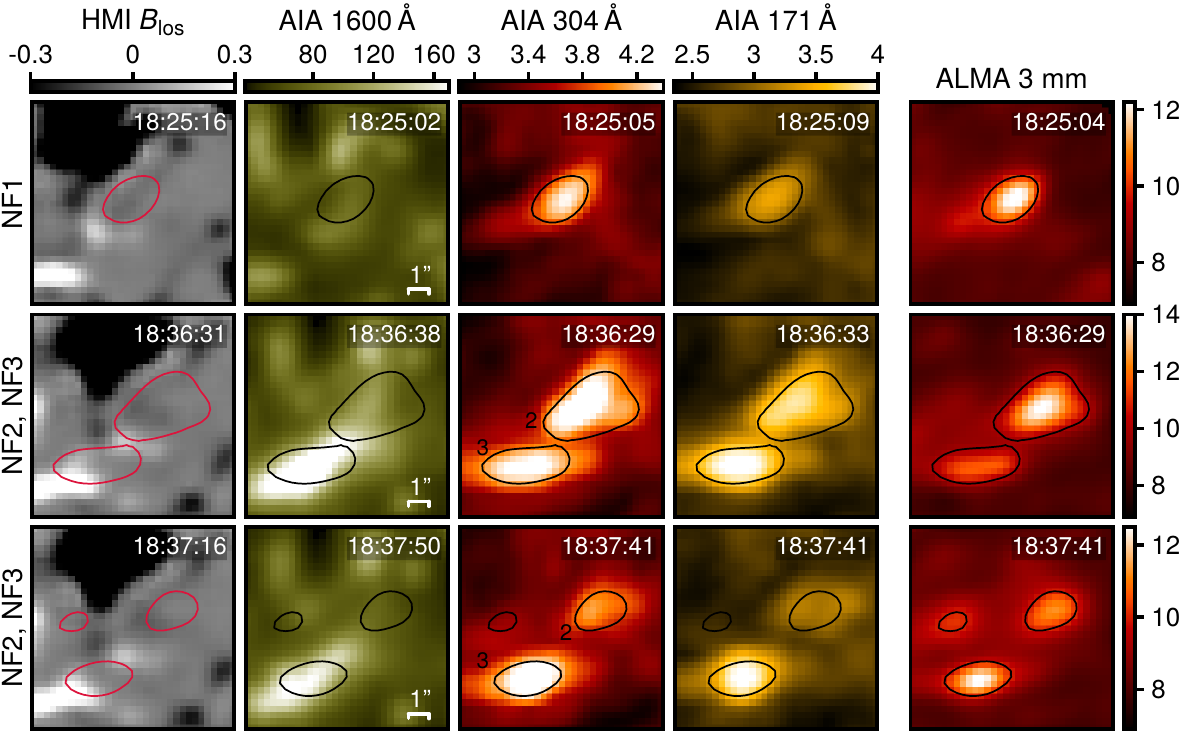}
    \caption{Examples of ALMA Band\,3 counterparts of EUV brightenings. From left to right: HMI magnetogram, spectral radiance in AIA\,1600\,\AA, AIA\,304\,\AA, and AIA\,171\,\AA, and ALMA 3\,mm brightness temperature.  All panels show a 10\arcsec$\times$10\arcsec~FOV centered on NFs. The contours correspond to $T_{\rm b}(3\,\mathrm{mm})=10$\,kK. The image scale is indicated in the panels in the second column. 
    The range in the HMI magnetograms is clipped at $\pm0.3$\,kG and the AIA images are capped at 170\,$\rm DN\,s^{-1}$ (1600\,\AA), $2.5\times10^{4}\,\rm DN\,s^{-1}$ (304\,\AA) and $10^{4}\,\rm DN\,s^{-1}$ (171\,\AA). The EUV images are displayed in logarithmic scale. The ALMA color bars are in units of kilokelvin.}
    \label{fig:mfpanel}
\end{figure*}

In the absence of H$\alpha$ observations, the 1700\,Å continuum can be used as a proxy for EBs \citep[e.g.,][]{2013ApJ...774...32V,2017GSL.....4...30C,2017ApJS..229....5D}. We searched for EB signatures in AIA\,1700\,\AA\ using the \texttt{EBDETECT}\footnote{\url{https://github.com/grviss/ebdetect}} code \citep{2019A&A...626A...4V}. The algorithm works by selecting events that reach a given intensity threshold above the QS mean in at least one 0.6" AIA\,1700\,\AA~pixel in two consecutive time frames. Here, we only consider $9\sigma$ events; this may sacrifice the recovery of some of the lower energy EBs, but it ensures that a higher fraction of the detections correspond to H$\alpha$ EBs. Given the well-known flickering property of EBs, we consider events that occur at the same location to be different if there is a considerable dimming over a period of several minutes between the adopted intensity threshold.
We identified a total of 20 EBs within the ALMA FOV throughout the time span of the observations.

Figure\,\ref{fig:ebpanel} shows three examples of EB detections (EB1, EB2, and EB3) in AIA and their corresponding visibility (or lack thereof) in ALMA at different times. The other events are displayed in the supplementary Fig.\,\ref{fig:ebpanelextra}.
The EBs appear as compact ($\lesssim$3\arcsec) sources in AIA\,1600\,\AA\ and 1700\,\AA,\ with lifetimes ranging from tens of seconds to several minutes, and at least half of the candidates are clearly associated with the interaction of opposite polarities (e.g., EB1, EB2, and EB3) as far as we can tell from the deconvolved HMI magnetograms. They usually have no counterpart in the EUV filters, or the emission seems unrelated to the UV brightenings. Likewise, we find that the EBs (both the $9\sigma$ cores and $5\sigma$ extended halos) are undetected in the ALMA Band\,3 maps.

In fact, EB2 is a striking example of how the 3\,mm continuum is not sensing the energy release of the EB, but it must be formed much higher in the chromospheric canopy. The panels show a strong enhancement in the UV continuum associated with a flux cancellation site that appears dark at 3\,mm with $T_{\rm b}$$\sim$7500\,K (similar to QS level), yet right next to it we clearly see a warm loop-like structure connecting the two main polarities' patches in the region. 
Events such as EB3 that occur within an extensive region that is generally enhanced in Band\,3 are inconclusive. In this case, the average $T_{\rm b}$ (and standard deviation) within the \texttt{EBDETECT} 5$\sigma$ contours is $\approx$8200$(\pm 350)$, but the lack of contrast between the EBs and the periphery does not allow us to unambiguously associated the 1700\,\AA~continuum enhancement with the 3\,mm brightness. 

We find that brightness temperature of the events located further way from the AR center were lower (down to QS levels) than the EBs occurring near the pores, which strongly suggests that enhanced $T_{\rm b}$ values are probably the result of the more "space-filling," persistent heating that is found in flux emerging regions \citep[e.g.,][]{2018A&A...612A..28L}, rather than a consequence of episodic chromospheric heating by EBs. Overall, the mean and standard deviation of the brightness of the EBs in Band\,3 are $T_{\rm b}$$\sim$7900$(\pm 300)\rm \,K$.

Figure\,\ref{fig:eblc} shows the light curves for the AIA\,1700\,\AA~filter and ALMA Band\,3 for EB1, EB2 and EB3 shown in Fig.\,\ref{fig:ebpanel}. They confirm that the photospheric 1700\,\AA~continuum enhancements and the overlying chromospheric mm continuum are weakly correlated in the EB candidates. After interpolating the mm observations to the AIA temporal sampling, the linear correlation coefficients are $r=$\,0.28, $r=$\,0.35, and $r=$\,-0.08 for EB1, EB2, and EB3, respectively. We also find no evidence for a lag between the photospheric and chromospheric signals. 

\subsection{The millimeter counterparts of EUV nanoflares}
\label{section:DEM}

Contrary to the EB events discussed in the previous section, we find nine bursty events in the AIA EUV images that have clear counterparts in the ALMA maps. These are also small ($\lesssim$4\arcsec) and short lived ($\sim$1--8\,min), but unlike the EB candidates that do not have a mm analogue, the EUV brightenings are usually accompanied by simultaneous (within the 12\,s cadence of the EUV data) mm continuum enhancement, or at least every time that there is a significant intensity increase in the 304\,\AA~channel. In most cases they only have a relatively weak signature in the 1600/1700\,\AA~filters compared to the plage region and EBs. We note that the ALMA data show other interesting transient brightenings that do not have an EUV analogue and could possibly be UVBs, but we did not quantify their occurrence in the same systematic way given the lack of context from IRIS.

In this paper, we focus on the two most significant events that occurred between 18:24--18:42\,UTC and feature three compact EUV kernels associated with bright loops. The kernels are some of the brightest features detected during the whole observation period in all AIA passbands, and we estimated their thermal energies to be within the nanoflare energy range (see Sect.\,\ref{section:contrib}).
Therefore, we labeled them as NF1 (first event) and NF2 and NF3 (second event). We adopted the nanoflare definition based on the energy estimate regardless of physical mechanism and expected occurrence rate \citep[e.g.,][]{2015RSPTA.37340256K}.

Figure\,\ref{fig:mfpanel} shows the three TR/coronal brightenings (NF1, NF2, and NF3) identified by visual inspection of the EUV data and their counterparts in the 3\,mm continuum. Other interesting brightenings are displayed in the supplementary Fig.\,\ref{fig:mfpanelextra}. NF1 appears as roundish bright source in all AIA EUV channels at the same location where we see an enhancement in the 3\,mm brightness temperature of the order of several thousand kelvin relative to the background. Interestingly, no significant 1600\,\AA~or 1700\,\AA~emission is detected. NF2 is a different event that occurs at the same location as NF1 approximately ten\,minutes later and near NF3. In the latter, there is a strong enhancement in the UV channels. 
The highest brightness temperature at 3\,mm that we observe in the EUV kernels is $T_{\rm b}$$\approx$14\,300\,K. In the three cases there is a remarkable spatial and temporal similarity between the EUV images and Band\,3 maps. We present the results of the analysis of the time series in Sect.\,\ref{section:faf}.

The insignificant 1700\,\AA~emission in NF1 but weak enhancement in the 1600\,\AA~filter (compared to average plage brightness) is indicative of a contribution from the \ion{C}{IV} doublet and other lines that may dominate the passband in flaring conditions \citep{2019ApJ...870..114S}. In the second event, we find a more pronounced brightening in both filters at NF3, which implies a mixed contribution from photospheric continuum and transition region lines and possibly suggests a different formation mechanism in the two events.

The average photospheric magnetic field strength is weak ($\lesssim$100\,G) at the flaring locations, and the association of NFs1-3 with opposite polarities in the photosphere is unclear and definitely not as obvious as in the EBs shown in Fig.\,\ref{fig:ebpanel}. We do not find clear evidence for magnetic flux cancellation as the driver of the three NFs. 

\subsection{Flaring fibrils}
\label{section:faf}

\begin{figure}
    \centering
    \includegraphics[width=0.93\linewidth]{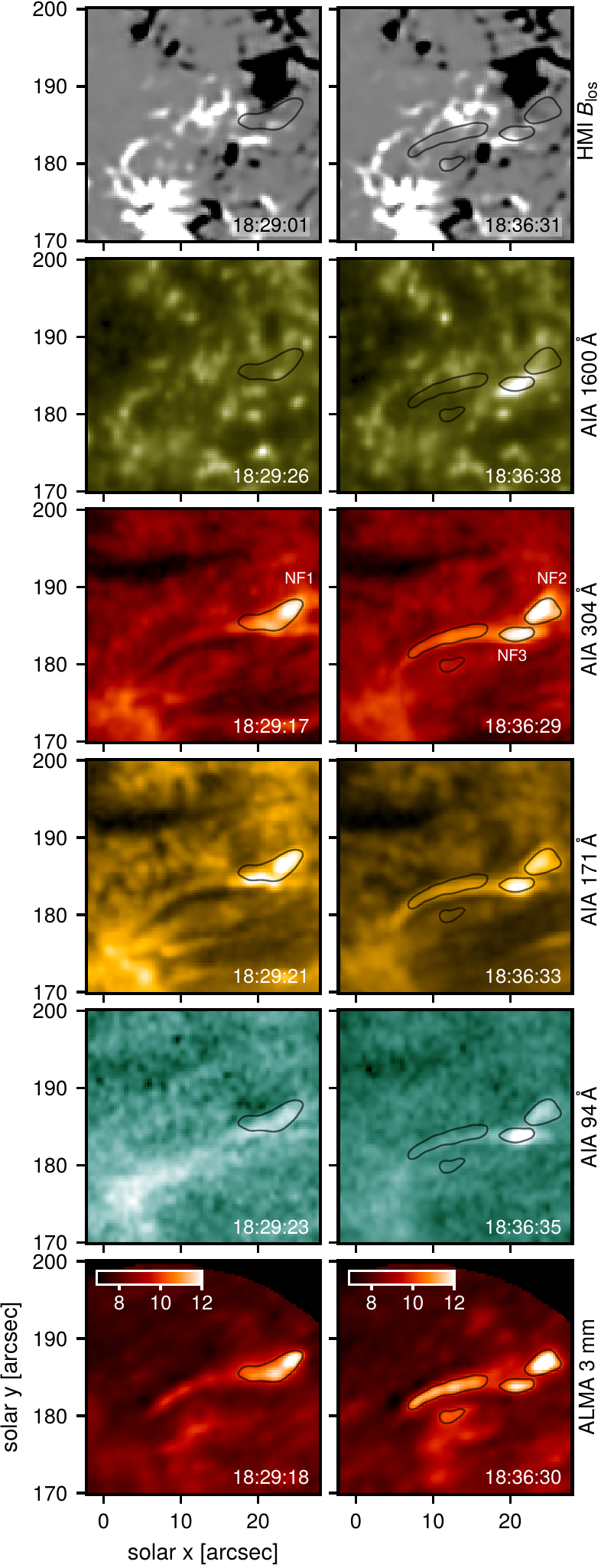}
    \caption{Flaring fibrils in SDO and ALMA. The range in the HMI magnetograms is clipped at $\pm0.3$\,kG. The AIA EUV images are displayed in logarithmic scale. The ALMA color bars are capped at 12\,kK for better visibility. The contours correspond to $T_{\rm b}(3\,\rm mm)=10$\,kK. An animated version of this figure showing the full time evolution is available \href{https://stockholmuniversity.box.com/s/oo9gkvu5q1q7dr7sxn8vq3xogk544bht}{online}.} \label{fig:fafevol}
\end{figure}

\begin{figure*}[h]
    \centering
    \includegraphics[width=\linewidth]{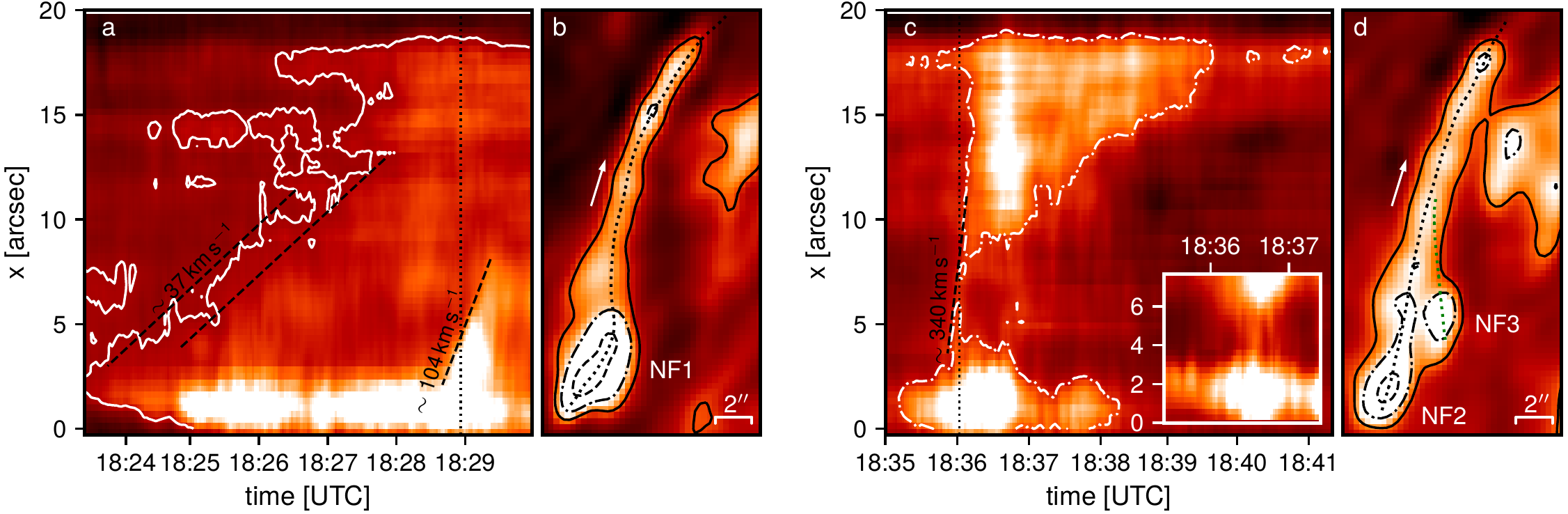}
    \caption{ Space-time diagrams for selected nanoflares and associated bright loops in the mm continuum. {\it Panels a} and {\it c} show space-time plots along the dotted lines in {\it Panels b} and {\it d} which in turn show ALMA $T_{\rm b}$ maps at two instances of time marked by the vertical dotted lines in the corresponding {\it panels a} and {\it c}. The contours correspond to 9, 10, 12\,kK (solid, dot-dashed, dashed). The inset in {\it panel c} shows a space-time plot for the green path in {\it panel d}. The space-time panels are saturated at 10\,kK. The dashed lines correspond to projected velocities of $37\rm \,km\,s^{-1}$ and $104\rm \,km\,s^{-1}$ in the first event, and $\sim$$340\rm \,km\,s^{-1}$ in the second event.}
    \label{fig:spacetime}
\end{figure*}

In this section, we analyze the temporal evolution of the two main events that feature the three NFs presented in Sect.\,\ref{section:DEM}. Inspection of the AIA data shows that NF1, NF2, and NF3 are associated with significant EUV loop brightenings akin to FAF events, which are also clearly visible in the ALMA maps. 

Figure\,\ref{fig:fafevol} shows selected instances of the evolution of the two events. An \href{https://stockholmuniversity.box.com/s/oo9gkvu5q1q7dr7sxn8vq3xogk544bht}{online} movie showing the full temporal evolution is also available. 
NF1, NF2, and NF3 occur in the vicinity of a particular fibril (e.g., Fig\,\ref{Fig:1} between $y=$\,180\arcsec--190\arcsec) that is about $\sim$18\arcsec~long and $\sim$1.5\arcsec~wide in the 3\,mm maps, and it is visible throughout the entire ALMA sequence. The brightness temperatures of the fibril are typically within the range of  8000--9000\,K, but reach much higher values (up to $\approx$12\,200\,K) in connection with the bright kernels.
The Band\,3 fibril is cospatial with an absorption feature in the AIA 171\,\AA, 193\,\AA~and 304\,\AA~channels between the two main opposite polarity patches, although it is not clear whether they are related. A filament is also visible in the AIA images to the north and away from the flux emergence region, but it appears practically indistinguishable from the local background at 3\,mm. 

In the first event that starts around 18:24 UTC, NF1 intermittently powers a faint EUV loop to which it is connected. The loop is best seen in AIA 171\,\AA~and 304\,\AA~but it is practically undetectable in AIA 94\,\AA~and 131\,\AA. NF1 eventually releases a fast-moving blob with an average FWHM of 1.9\arcsec$\times$2.8\arcsec\, as measured in the AIA 171\,\AA, 193\,\AA~and 304\,\AA~filters that have higher signal-to-noise ratio. 
The same temporal evolution is echoed in the ALMA Band\,3 maps. The loop is initially between $T_{\rm b}$$\approx$8000--8500\,K, but it gradually warms up to $T_{\rm b}$$\approx$9000--9500\,K from a stream of plasma that travels with a projected velocity of approximately $\approx$$37\rm\,km\,s^{-1}$ from the NF kernel as the latter oscillates in brightness (see Fig.\ref{fig:spacetime}). Analogous flickering intensities and similar velocities were observed in high-resolution SST \ion{Ca}{II}\,K filtergrams of plasmoid-like fine structure in UVBs \citep{2017ApJ...851L...6R}. The blob is also visible in Band\,3 and resembles the plasmoid ejection described in \citet{2017ApJ...841L...5S} but shows higher temperatures ($T_{\rm b}$$\approx$$12\,400$\,K) and EUV intensities. 

The second event starts at around 18:35 UTC in the same area of the first event, but it features two different EUV brightenings (NF2 and NF3) and a more pronounced flaring fibril that is partially visible in AIA 94\,\AA.
The loop reaches $T_{\rm b}\sim12\,200$\,K at 3\,mm more impulsively than in the first event, and we detect a much faster heat front with $T_{\rm b}$$\approx$10\,000\,K and a plane-of-sky velocity of $\approx$$340\rm\,km\,s^{-1}$ (see Fig.\ref{fig:spacetime}), which is best seen at the high cadence of the ALMA observations. 

Interestingly, we do not observe any cospatial 1600\,\AA~emission in the hot loops in either of the events, unlike, for example, in the FAFs reported in \citet{2015ApJ...812...11V} or the transient brightenings in \citet{2009ApJ...701.1911P} and \citet{2018ApJ...857..137G}, which suggests that the bulk of the plasma is not at temperatures between $\log T$$\sim$4.2-5.1\,K to which AIA 1600\,\AA~would be sensitive \citep{2019ApJ...870..114S}. 

\begin{figure*}[h]
    \centering
    \includegraphics[width=\linewidth]{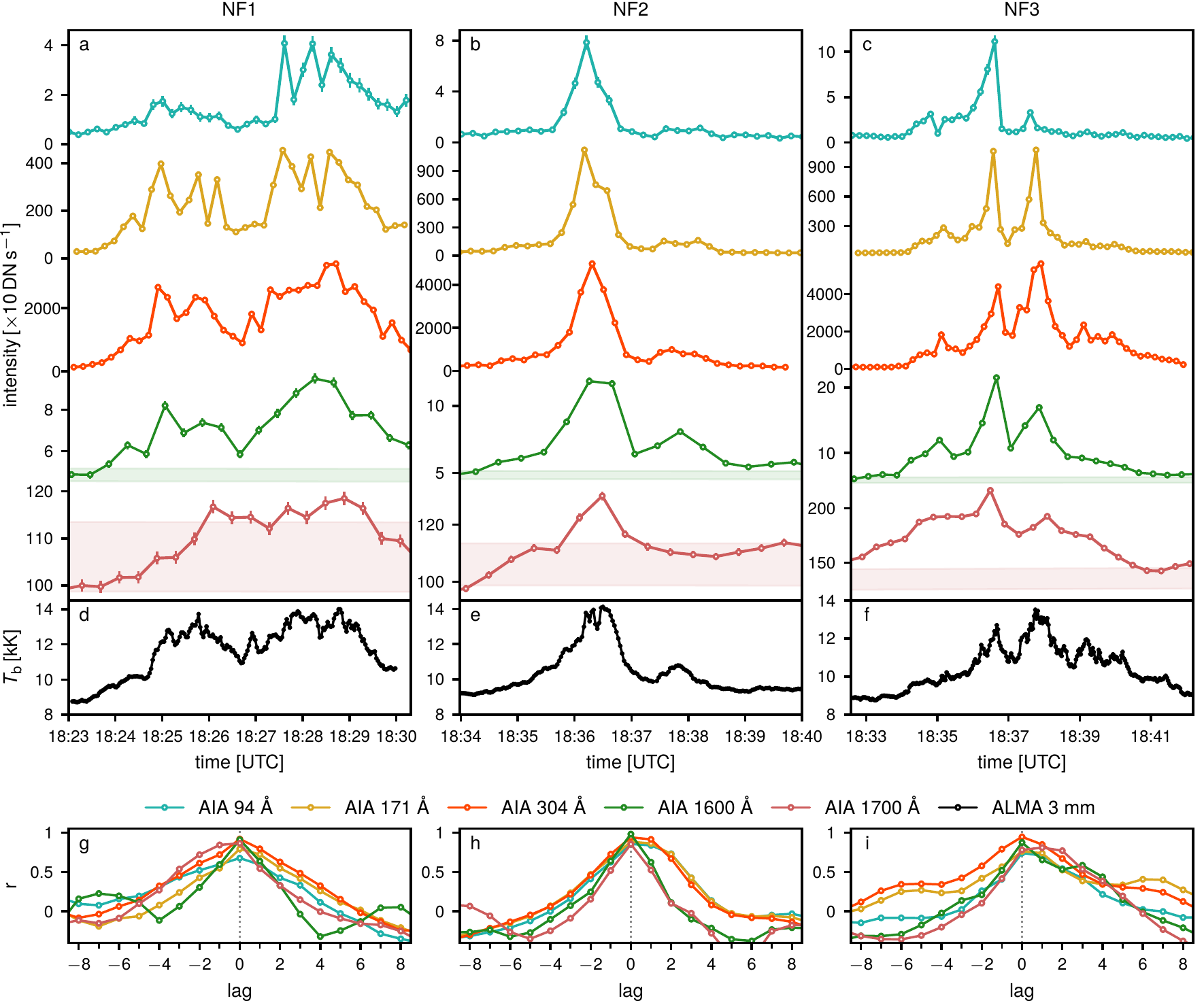}
    \caption{ AIA and ALMA light curves. {\it Panels a, b,} and {\it c}: intensity as function of time in different AIA passbands at the three nanoflares shown in Fig.\,\ref{fig:mfpanel}; the shaded areas in the 1600\,\AA~and 1700\,\AA~panels enclose the mean\,$\pm1\sigma$ of the (detrended) signal over a longer period of time. {\it Panels d, e,} and {\it f}: the corresponding mm brightness temperature. {\it Panels g, h} and {\it i}: cross-correlation functions between ALMA and AIA; one lag unit corresponds to 12\,s in the EUV and 24\,s in the UV.} \label{fig:lcurves}
\end{figure*}

Figure\,\ref{fig:lcurves} shows the AIA and ALMA light curves at the center of the three NFs discussed above and the respective cross-correlation obtained for different temporal displacements (or lag) of the ALMA signal relative to AIA. The latter are computed by interpolating the ALMA observations to the AIA cadence. We find that there is not only spatial correspondence but also temporal coherence between the EUV and mm emission since the correlation values approach unity and are maximal or nearly flat-topped at lag zero, which means that the brightenings are nearly simultaneous in all wavelengths or have a relative delay that is of the order of the AIA cadence at most.

We investigated if there are any periodic components in the more long-lasting NF signals using wavelet analysis and found periodicities in the range $\sim$50-150\,s (see Appendix\,\ref{section:oscillations}). Similar time scales were identified in Interface Region Imaging Spectrograph \citep[IRIS, ][]{2014SoPh..289.2733D} FUV observations of explosive events and linked to variations in the photospheric magnetic field strength \citep{2015ApJ...809...82G}. We do not find evidence for the latter in the events presented here. 

\subsection{Differential emission measure analysis}
\label{section:contrib}

\begin{figure}
    \centering
    \includegraphics[width=\linewidth]{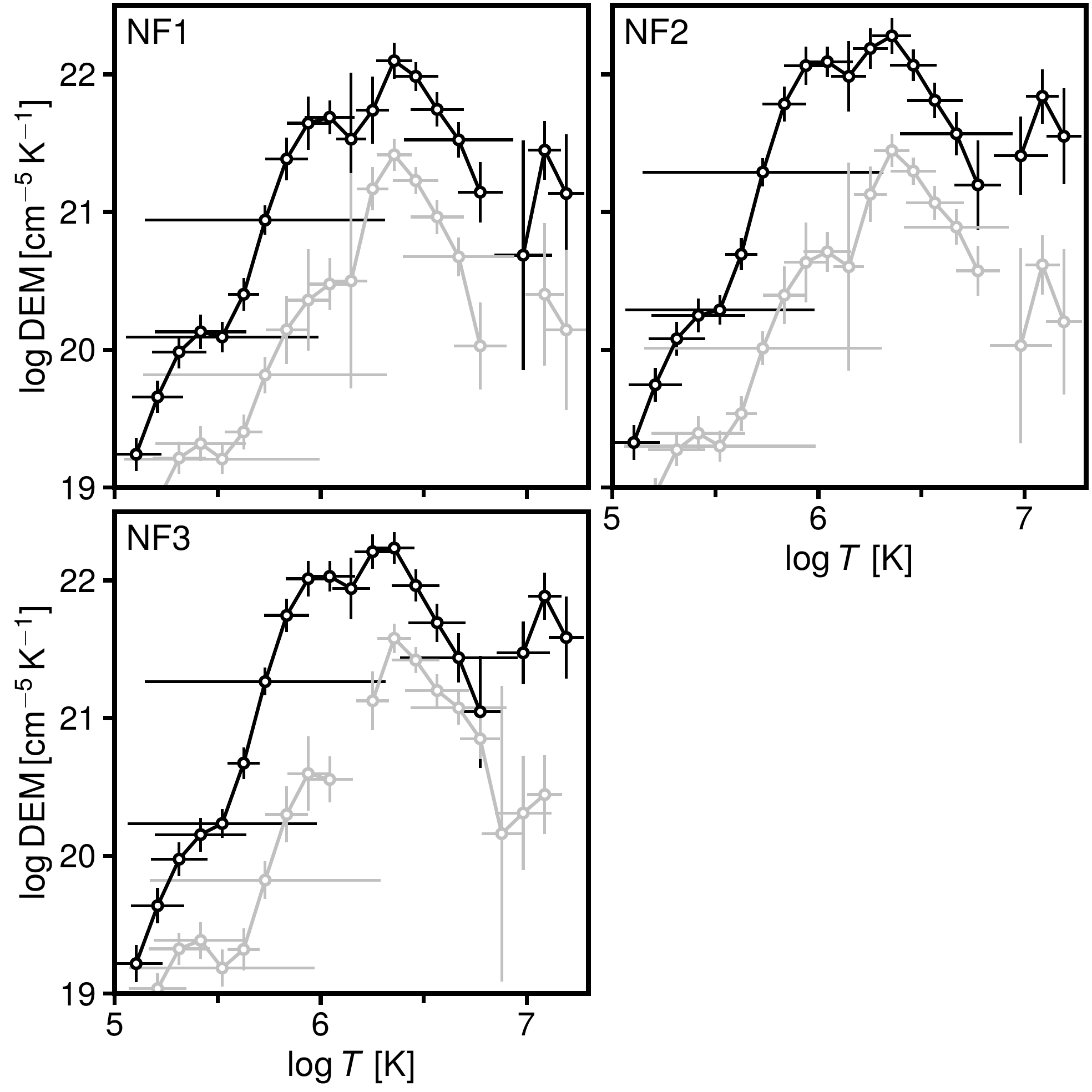}
    \caption{ Differential emission measure analysis on AIA EUV data. DEM curves at peak brightness (black) and respective backgrounds (gray) for the events NF1, NF2, and NF3.}
    \label{fig:demplots}
\end{figure}

\begin{figure*}
    \centering
    \sidecaption
    \includegraphics[width=120mm]{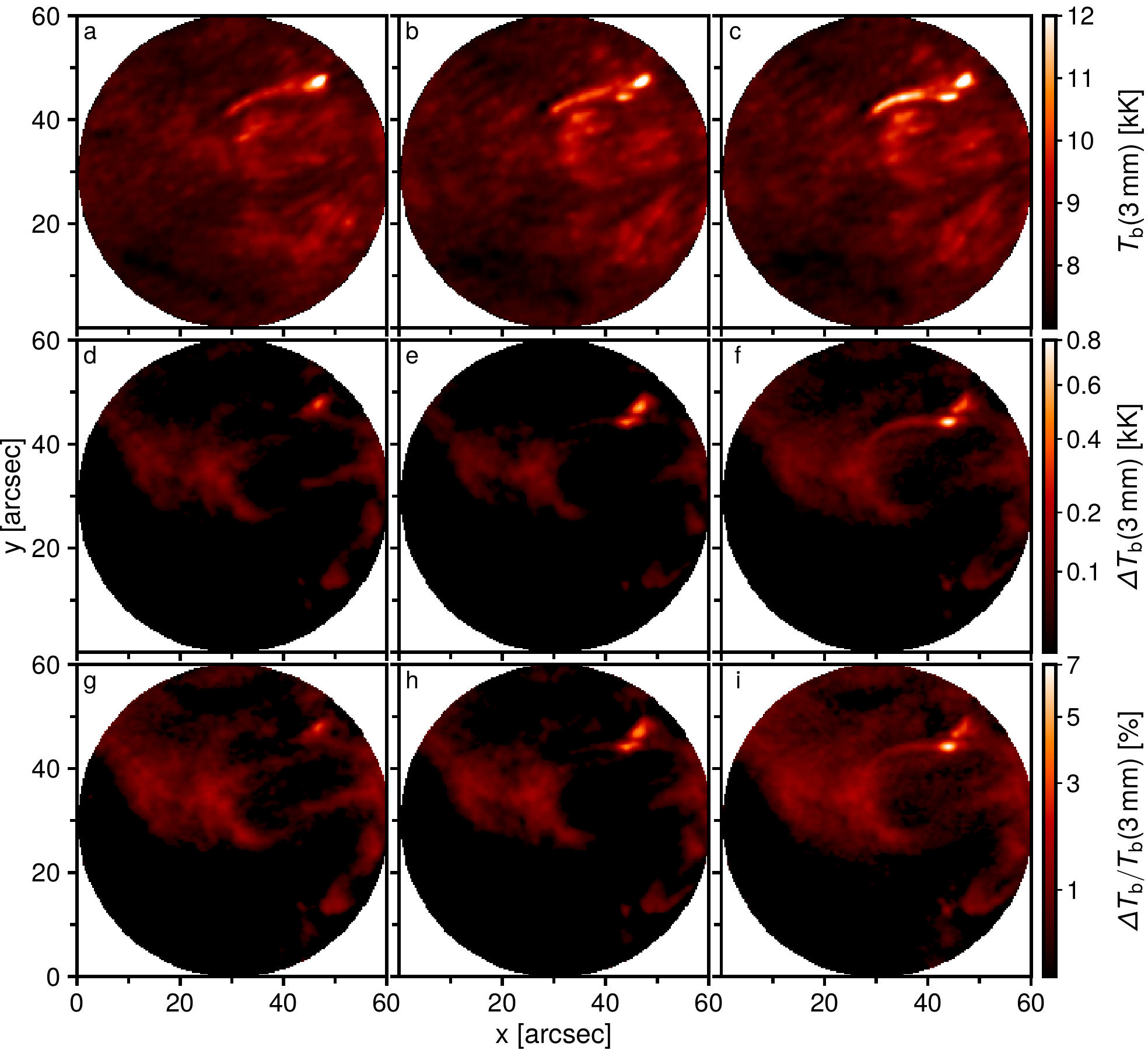}
    \caption{Coronal contribution to the 3\,mm brightness temperature in the active region. {\it Panels a-c:} 10-second average of the observed $T_{\rm b}$ at 3\,mm (capped at 12\,kK) at different times (18:28:45 UTC left, 18:36:15 UTC middle, and 18:36:35 UTC right). {\it Panels d-f:} contribution from coronal plasma to $T_{\rm b}$ at 3\,mm. {\it Panels g-i:} relative contribution in percentage. {\it Panels d-i} are shown in power-law scale.}
    \label{fig:contrib}
\end{figure*}

We estimated an order of magnitude for the thermal energy of the EUV kernels (Sect.\,\ref{section:DEM}) as $E_{\rm th}$$\sim$3$n_{\rm e}k_{\rm B}TV,$ where $T$ is the electron temperature, $n_{\rm e}$ is the electron density, and $V$ is the volume that is approximated as $l\,w^2$ for brightenings with a certain projected length $l$ and width $w$ on the disk \citep[e.g.,][]{2004psci.book.....A}. The dimensions are estimated by fitting a 2D Gaussian to the kernels and computing the full width at half maximum (FWHM) along the orthogonal axes, while the depth is assumed to be the same as $w$. The temperature is obtained from the peak of the differential emission measure (DEM) curve that is inferred from the AIA 94\,\AA, 131\,\AA, 171\,\AA, 193\,\AA, 211\,\AA, and 335\,\AA~filters using the regularized DEM inversion code\footnote{\url{http://www.astro.gla.ac.uk/~iain/demreg}} described in \citet{2012A&A...539A.146H}. 
We use time-dependent AIA instrumental responses obtained from SSW with CHIANTI v9 \citep{1997A&AS..125..149D,2019ApJS..241...22D} to account for instrumental degradation on the uncorrected count rates, and we include a systematic uncertainty factor of 20\% to account for uncertainties in atomic data.

Figure\,\ref{fig:demplots} shows examples of the DEM inversions of spatially-averaged intensities (4$\times$4\,px, 1.2\,arcsec${}^2$) at the center of NF1, NF2, and NF3 at peak brightness. 
We note that the inverted DEM curves are only reliable from $\log T$$\gtrsim$5.5\,K and some large errorbars are related to sensitivity gaps in the AIA response functions. 
We find an increase of the DEM at all temperature bins compared to the background. Despite the greatest relative increase being at $\approx$$10^{5.9}$\,K, the highest peak of the DEM curves is at $\approx$$10^{6.4(\pm0.1)}$\,K, even after background subtraction. The background intensities are defined as the values just before the onset of the events at the same locations. 

Assuming a typical active region coronal density of $n_{\rm e}$$\sim$$10^9\rm\,cm^{-3}$ and an estimated size of $\approx$1.4\arcsec$\times2.6$\arcsec (NF1) and $\approx$1.5\arcsec$\times2.7$\arcsec (NF2 and NF3) we find $E_{\rm th}$$\approx$$2.1(\pm0.5)\times10^{24}$\,erg (NF1) and $E_{\rm th}$$\approx$$2.3(\pm0.5)\times10^{24}$\,erg (NF2 and NF3).
If we estimate the electron densities using $n_{\rm e}$$\sim$$\sqrt{EM/w}$, where $EM=\int \mathrm{DEM}(T)\,dT$ is the plasma column emission measure, and a filling factor of one is assumed, we find $n_{\rm e}$$\approx$$2\times10^{10}\rm\,cm^{-3}$ (NF1) and $n_{\rm e}$$\approx$$2.7\times10^{10}\rm\,cm^{-3}$ (NF2 and NF3). Therefore, the energies are $E_{\rm th}$$\approx$$4(\pm1)\times10^{25}$\,erg (NF1) and $E_{\rm th}$$\approx$$6(\pm1)\times10^{25}$\,erg (NF2 and NF3).
These estimates are within the typical NF energy range $\sim$$10^{24}$--$10^{27}$\,erg \citep[e.g.,][]{2004psci.book.....A} and similar values have been inferred from observations of small-scale EUV brightenings \citep[e.g.,][]{2000ApJ...535.1047A,2000ApJ...529..554P,10.1093/mnras/sty1712}.

The spatio-temporal correspondence between the bright EUV kernels and the mm bursts raises the question whether there is a significant contribution from the corona to the brightness temperatures in Band\,3. 
Following the approach of \citet{2013PASJ...65S...8A} and \citet{2017ApJ...845L..19B}, the contribution to the mm intensity along the $z$-direction on the line of sight from the optically thin corona can be estimated as follows:
\begin{equation}
    \Delta I_{\rm \nu} \approx \int \alpha_\nu S_\nu dz
    \label{eq:contrib}
,\end{equation}
\noindent where $S_{\nu}=B_{\nu}(T)$ is the source function that is equal to the Planck function, and $\alpha_{\nu}$ is the free-free absorption coefficient that is given by:
\begin{equation}
    \alpha_{\nu} = \frac{4e^6}{3hc}\left(\frac{2\pi}{3km^3_{e}}\right)^{1/2} \frac{ n_{e}}{T^{1/2}\nu^3} \sum_{i} Z^2_{i}n_{i}\,g_{\rm ff}(\nu, T) (1-e^{-h\nu/kT})
,\end{equation}
\noindent where $h$ is the Planck constant, $e$ is the electron charge, $c$ is the speed of light, $k$ is Boltzman constant, $m_{\rm e}$ is the electron mass, $\nu$ is the frequency, $Z_{i}$ and $n_{i}$ are the charge number and density of the ion species $i$, $ g_{\rm ff}$ is the Gaunt factor of free-free processes, and the last term in parentheses is the correction for stimulated emission.
In the Rayleigh-Jeans limit, the expression above can be approximated as given by \citet{1985ARA&A..23..169D} in cgs units:
\begin{equation}
    \alpha_{\nu}\approx 9.78\times10^{-3}\frac{n_{\rm e}}{\nu^{2}T^{3/2}}\sum_{i}Z_{i}^{2}n_{i}\,(24.5 + \ln T - \ln \nu) \label{eq:opacity}
,\end{equation}
\noindent for $T>2\times10^{5}$\,K. Substituting Eq.\,\ref{eq:opacity} in Eq.\,\ref{eq:contrib} and assuming a fully ionized hydrogen plasma, the contribution defined in terms of brightness temperature is given as follows:
\begin{equation}
    \Delta T_{\rm b}\approx 9.78\times10^{-3} \nu^{-2}\int (24.5+\ln T + \ln \nu )T^{-1/2} n_{\rm e}^2 dz
    \label{eq:contrib1}
.\end{equation}
In reality, the stratification of electron density with height is not known, but we can use the estimated DEM curves instead from the definition:
\begin{equation}
    \mathrm{DEM}(T)=n_{\rm e}^{2}\frac{dz}{dT} 
.\end{equation}
This means one expects a larger contribution from the corona in the heating events where there is a significant enhancement of the plasma emission measure (see Fig.\,\ref{fig:demplots}). We computed the DEM in each pixel of the FOV, and the integration was carried between 5.5$\leq\log T\leq$7.3 within the range of sensitivity of the AIA bands. 

Figure\,\ref{fig:contrib} shows examples of the contribution from the corona to the observed 3\,mm brightness temperature for three instances of the main events discussed in this paper. We performed a ten-second average on the ALMA data to account for the different time sampling of each AIA channel.
We find that the contribution from the corona in Band\,3 is $\lesssim$1\% ($\lesssim$73\,K) in the quieter areas, but it increases to several hundred kelvin at the NF sites. We find a relative contribution of $\sim$3\% in NF1, 4\% in NF2, and 7\% in NF3.

This suggests that a non-negligible (in the sense that it is larger than the noise root-mean-square) contribution from the corona may overlap with the chromospheric signal detected with Band\,3 in NF events, or more generally in flaring conditions in high density media. The mm brightness may remain significantly enhanced for several minutes after the strong emission in the hotter AIA channels has vanished as shown on the right panel in Fig.\,\ref{fig:lcurves}, but it still correlates with the intensity enhancements detected in the 304\,\AA~filter. The contributions in the bright loops are lower (1--2\%). 

The aforementioned estimates can be considered as lower limits since we did not assess the contribution of the transition region ($\log T$<5.5\,K) to the mm brightness, which is in principle more important than the corona given the $T^{-1/2}$ dependence in Eq.\,\ref{eq:contrib1}. 
We note that these results are subject to the uncertainties in the flux calibration of both instruments and uncertainties inherent to the DEM inversion due to the limited information that can be extracted from the response functions of the EUV channels alone. 

\section{Synthetic millimeter continuum from a 3D flux emergence simulation}
\label{section:simulation}

\begin{figure*}
    \centering
    \includegraphics[width=\linewidth]{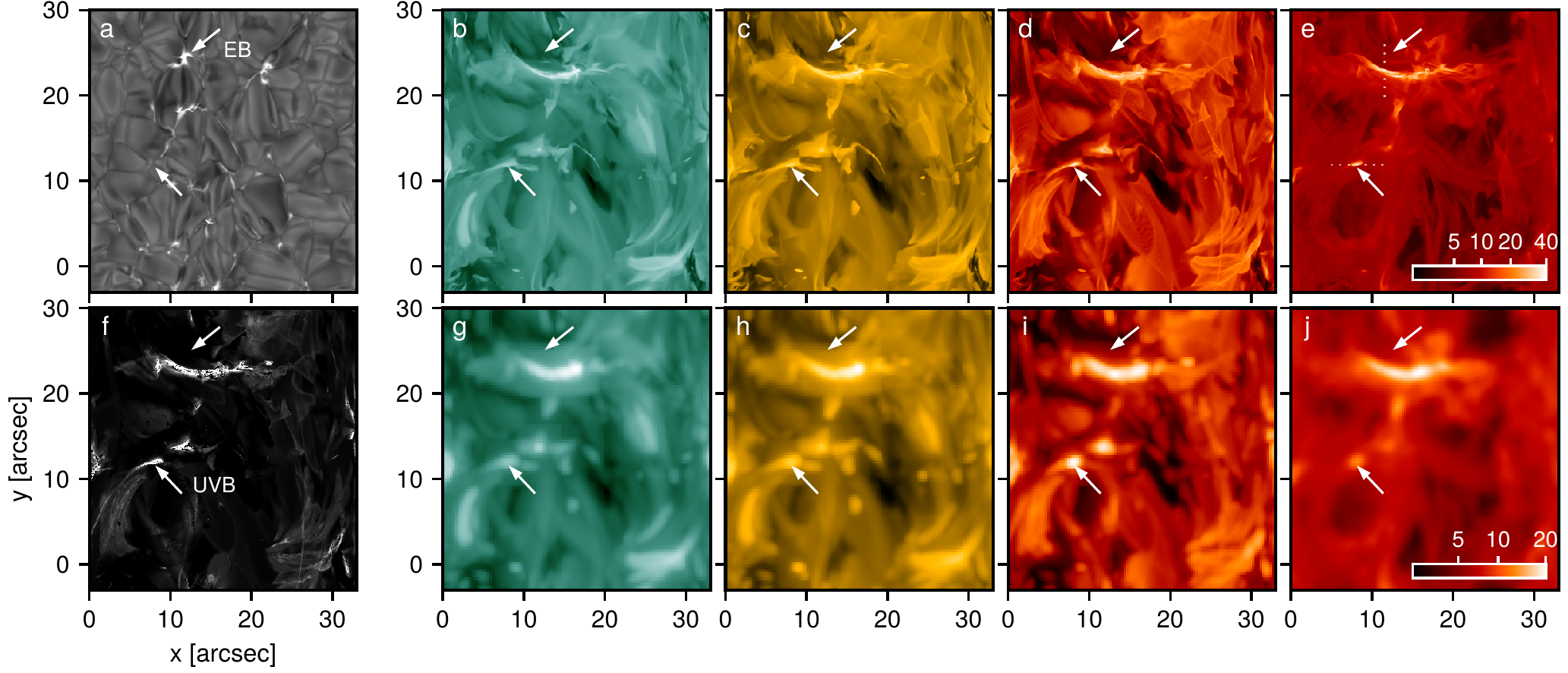}
    \caption{Synthetic emission at different wavelengths from a 3D r-MHD simulation. For context, {\it panels a} and {\it f} show the H$\alpha$ line wing (-1.5\,\AA)~and integrated \ion{Si}{IV}\,1393\,\AA~intensity, respectively. The emission at the simulation's original scale ($0.065\arcsec$/px) in AIA\,94, 171, 304\,\AA,~and ALMA 3\,mm is shown in {\it panels b, c, d,} and {\it e}, respectively, and at the spatial resolution of the observations ($0.3\arcsec$/px) in {\it panels g, h, i,} and {\it j}, correspondingly. The arrows indicate the locations of one EB and one UVB as labeled in the leftmost panels. {\it Panels b-d, g-i} are displayed in logarithmic scale; {\it panels e, f, j} are shown in power-law scale. The dotted lines in the upper-right panel indicate the locations of the slices shown in Fig.~\ref{fig:taus}. } 
    \label{fig:simulationOld}
\end{figure*}

\begin{figure}
    \centering
    \includegraphics[width=\linewidth]{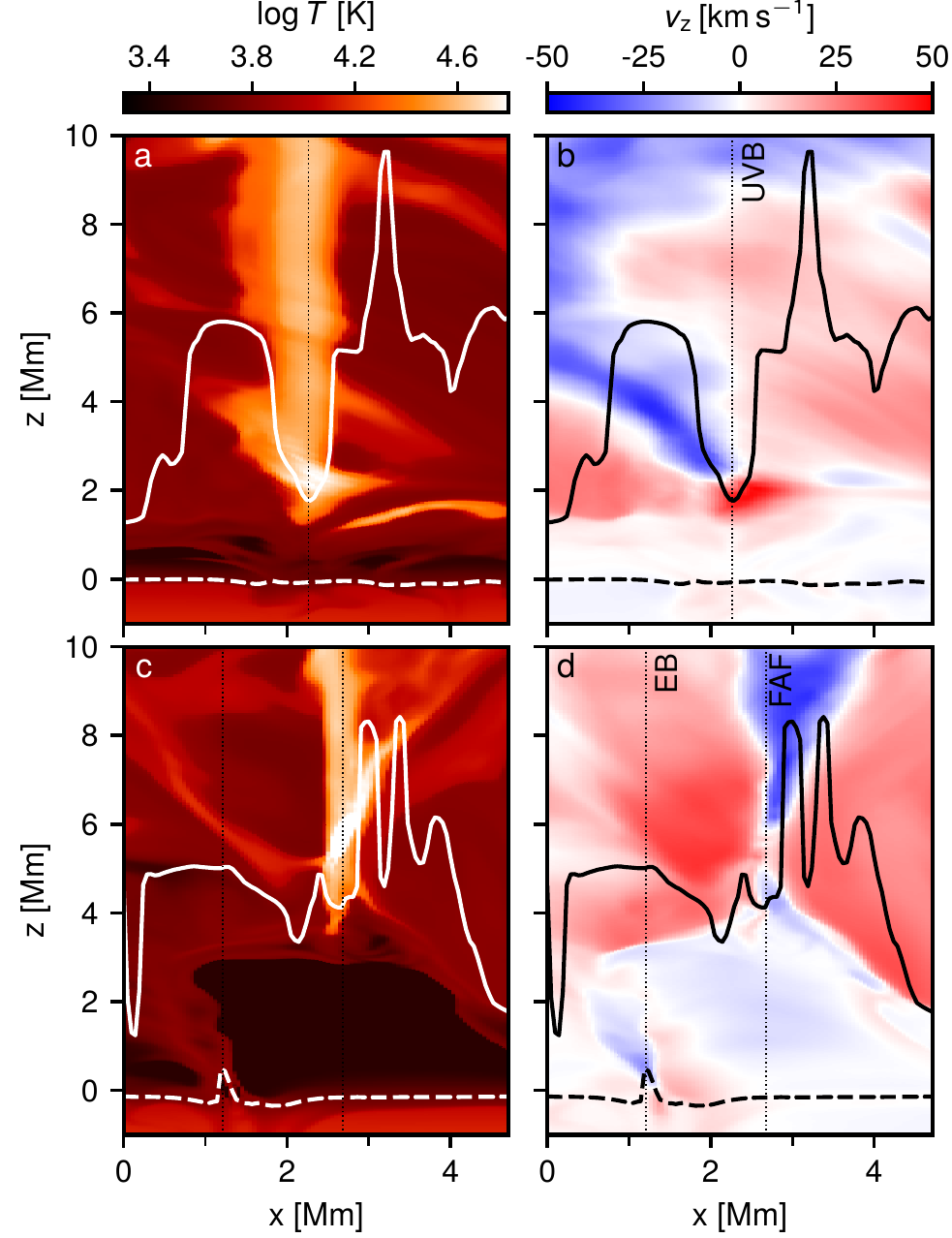}
    \caption{ Vertical cuts through the simulation box at the location of reconnection events. Top row: cut through the UV burst; bottom row: cut through the EB and a FAF. The left-hand column shows the temperature, the right-hand column the vertical velocity. The $\tau=1$ layer of H$\alpha$-1.5\,\AA~(dashed) and 3\,mm (solid) radiation are shown in all panels. The thin dotted lines indicate the locations of the EB, UVB, and FAF.}
    \label{fig:taus}
\end{figure}

In order to address the broader question as to whether ALMA Band\,3 is capable of detecting EBs, UVBs, and similar small-scale heating events, we used one snapshot of a 3D r-MHD flux emergence simulation performed with the \texttt{Bifrost} code \citep{2011A&A...531A.154G} and analyzed by \citet{2017ApJ...839...22H}, to which we refer for a detailed description of the setup. The box size is 504$\times$504$\times$496 px and the resolution is 48\,km per px in the horizontal direction and between 20-100\,km per px in the vertical direction from the photosphere to the corona.

We complement the synthetic visible and UV diagnostic images of \citet{2017ApJ...839...22H} with 3\,mm thermal continua that were computed in nonlocal thermodynamic equilibrium (NLTE) using the STockholm Inversion Code\footnote{\url{https://github.com/jaimedelacruz/stic}} \citep[\texttt{STiC},][]{2016ApJ...830L..30D,2019A&A...623A..74D}.
The calculations are carried out iteratively by solving the statistical equilibrium equations for a six-level hydrogen atom imposing charge conservation so that the electron densities $n_{\rm e}$ are consistent with the NLTE hydrogen populations. The ionization balance of other atoms is treated in LTE in the equation of state.

In addition, we computed the optically thin emission in the AIA EUV channels using version 8 of the temperature response functions $K_{\mathrm{ch}}(T)$ \citep{2014SoPh..289.2377B} that are obtained through the \texttt{aia\_get\_response} routine in SSW using standard coronal abundances and assuming a fully ionized plasma, such that the intensity $I$ in each channel can be calculated as given below:
\begin{equation}
    I_{\rm ch}= \int n_{\rm e}^{2}(z)K_{\mathrm{ch}}(T)dz.
\end{equation}

Figure\,\ref{fig:simulationOld} shows synthetic observables computed from the \texttt{Bifrost} simulation. Synthetic H$\alpha$ and \ion{Si}{IV}\,1393\,\AA~images are shown for identification of EBs and UVBs that were previously discussed and compared with observations in \citet{2017ApJ...839...22H}. 
The average $T_{\rm b}$ at 3\,mm in the whole box is approximately $\approx$6700\,K, which is lower than the observed QS average \citep[$\approx$7300\,K,][]{2017SoPh..292...88W}. Furthermore, the magnetic topology is not directly comparable to the observed active region, so we cannot assess whether the observed events originate in the same way as in the simulation. However, the synthesized mm continua shows that the flux emergence process is able to significantly heat up the chromosphere in different reconnection events.
We find a good correspondence between $T_{\rm b}$ at 3\,mm and the \ion{Si}{IV} brightenings, but no counterparts to the H$\alpha$ EBs. This is due to the deeper formation of the EBs relative to the UVBs and small flares in the simulation (see below). We note that the analysis in this paper does not allow us to conclude that reconnection is the cause of the observed Band\,3 brightenings, although it is plausible that some of the events are caused by the interaction of the emerging magnetic loops and the preexistent field as in the simulation.

The highlighted EB in Fig.\,\ref{fig:simulationOld} has an average $T_{\rm b}$ of $\sim$6800\,K, whereas the UVB shows a range of temperatures between $\sim$20--37\,kK; the highest values are only reached in a few pixels, so they would not be resolved in our ALMA data. Smearing the synthetic data to the spatial resolution of the observations using a Gaussian kernel brings the maximum $T_{\rm b}$ down to $\sim$13\,kK, which agrees with the typical values that we observe at the bright kernels (see Fig.\,\ref{fig:mfpanel}). The simulation also produces a FAF-like feature in the mm continuum that is remarkably similar to the observations (see Fig.\ref{fig:fafevol}), although the maximum synthetic brightness of $\sim$19\,kK at 1.2\arcsec~resolution is much larger than the observed ($\sim$12\,kK). 

In this simulation, we also find a good correspondence between ALMA and the EUV brightenings, especially with the 304\,\AA~passband as in the observations, which can be understood from the temperature response of this band to relatively cooler ($T<10^5$\,K) plasma.
However, detailed optically thick, nonequilibrium radiative transfer is needed to more accurately model the \ion{He}{II}\,304\,\AA~line \citep[e.g.,][]{2014ApJ...784...30G,2017A&A...597A.102G} that dominates this passband.

Figure\,\ref{fig:taus} shows two cuts through the simulation along the horizontal (bottom panels) and vertical (lower panels) lines drawn in {\it panel e} in Fig.\,\ref{fig:simulationOld}. We plotted the layers where the optical depth of the 3\,mm continuum and the wing of H$\alpha$ at -1.5\,\AA~from the line center reach unity. 
The 3\,mm continuum is formed over a broad range of heights that spans several mega-meters due to the expansion of the chromosphere in this simulation due to the magnetic flux emergence process.  
At the reconnection sites, the $\tau_{3\rm\,mm}=1$ layer drops due to the large temperature increase in the overlying atmosphere ($\alpha_{\rm \nu}\propto T^{-3/2}$).
Figure\,\ref{fig:taus} also shows that the formation height of the EB-like brightening in the wing of H$\alpha$ is much lower than the 3\,mm radiation, which is consistent with the observational findings (see Sect.\,\ref{section:EBsandMFs}).

We find that the brightness temperature at 3\,mm and the gas temperature at $\tau_{\rm 3\,mm}=1$ are well-correlated ($r=0.92$), but these two quantities may differ by several thousand kelvin at certain locations, particularly in the upflowing jets, due to the integration of contributions from different layers including the TR.
At the center of the UVB and FAF, as indicated by the vertical dotted lines in Fig.\,\ref{fig:taus}, $T_{\rm b}(3\,\rm mm)$ provides a good direct estimate of the temperature of the reconnection.

In this simulation the contribution from the corona to the mm intensities is negligible, as most of it has been pushed away by the emerging magnetic fields. For example, in the flaring fibril the optically thin contribution from plasma at $T$$\geq$$10^5$\,K is approximately 1\% at 3\,mm. This may not be the case in real active regions on the Sun (see Sect. \ref{section:contrib}). 

\section{Discussion}
\label{section:discussion}

In the comprehensive study of \citet{Georgoulis_2002}, EBs were described as bright features in the wings of H$\alpha$ (and often in the 1600\,\AA\,continuum) with a typical size of 1.8\arcsec$\times$1.1\arcsec~and occurring at a rate of at least $1.43\rm \,min^{-1}$ in a FOV of 1800\,arcsec,${}^{2}$ and it is speculated they could be important for chromospheric and even coronal heating. EBs with such properties would readily be detected in our ALMA observations. However, not even the brightest and largest candidates that we found in the 1700\,\AA~continuum images had a mm counterpart, which is in line with the prediction of \citet{2017A&A...598A..89R}. 
 
This is consistent with observational and numerical studies that propose low-altitude reconnection as the causing mechanism of EBs \citep[e.g.,][]{Georgoulis_2002,2007A&A...473..279P,2008PASJ...60...95M,2009A&A...508.1469A,2011ApJ...736...71W,2013ApJ...774...32V,2017A&A...601A.122D,2017ApJ...839...22H,2019A&A...626A..33H}, and link temperature enhancements around the temperature minimum region with their common spectral signatures \citep[e.g.,][]{2006ApJ...643.1325F,2010MmSAI..81..646B,2013A&A...557A.102B,2015ApJ...810..145D,2015RAA....15.1513L,2017ApJ...835L..37R,2017ApJS..229....5D,2019A&A...627A.101V}. 
We note that ALMA Band\,3 observations might still detect EBs that have a UVB counterpart as in, for example, \citet{2015ApJ...812...11V,2016ApJ...824...96T,2017A&A...598A..33L}. Future ALMA observations should aim to search for EBs using Band\,7 as it probes lower heights closer to the classical temperature minimum. 

In contrast, we identified multiple events with brightness temperatures above 9\,kK in the ALMA sequence that coincide with bright EUV structures.
The thermal energies of the order of a few $\sim$$10^{24}$-$10^{25}$\,erg along with the lack of strong photospheric signal, suggest they are NF events occurring in higher layers of the atmosphere but the causing mechanism is unclear. Some of these mm-bursts could be UVBs, although it is uncommon for the latter to show strong EUV emission \citep[e.g.,][]{2018SSRv..214..120Y}, likely due to absorption by cool gas along the line of sight \citep{2019A&A...626A..33H} or because temperatures above $\sim$0.1\,MK may not be reached \citep{2019A&A...628A...8P}. 

It may be that the events described in this paper are of the same class as the peculiar UVB reported in \citet{2018ApJ...856..127G} that was fully visible in all AIA channels. Those authors argue that the magnetic topology was such that it allowed a magnetic reconnection event at higher heights than usual, hence the coronal emission. The aforementioned line-of-sight effects could also play a role, with insufficient absorbing cool gas to obscure the events from view.

\citet{2014Sci...346C.315P} originally proposed that UVBs occur in the photosphere, while magnetic field extrapolations place the reconnection site in the low chromosphere $\sim$500-1000\,km \citep{2017A&A...605A..49C,2018ApJ...854..174T}. In these circumstances it is not clear whether UVBs would be as obscured as Ellerman bombs by the chromospheric canopy if the mm continuum in active regions is formed at much higher heights than previously thought \citep{2020ApJ...891L...8M}. 
In the simulation of \citet{2017ApJ...839...22H} the UVBs occur at chromospheric densities and originate strong mm emission (Sect.\,\ref{section:simulation}). The increase of ionization degree in the chromosphere will further raise the electron-proton free-free opacity of the mm continuum, ensuring that $T_{\rm b}$ is a good proxy for the plasma temperature. 
IRIS observations are needed to definitively confirm whether the mm-bursts have UVB signatures as this would impose important constraints on their formation height.

We investigated the contribution of the corona to the mm brightenings from DEM analysis using six AIA channels and found that there could be a significant contribution between $\sim$4-7\% from plasma at $T>10^{5.5}$\,K in different NF events, although this only partly explains the enhancements of up to $\sim$60\% in Band\,3 relative to the background.
The cooler TR plasma is also expected to contribute to the observed brightness. Therefore, the emission detected by ALMA Band\,3 in small (and larger) flares may result from a contribution from a broad distribution of plasma temperatures. The \texttt{Bifrost} simulation shows $T_{\rm b}$ at 3\,mm around $\sim$37\,kK at spatial scales much smaller than what ALMA can resolve at the moment.
Everywhere else in the AR, including the bright loops and periphery, the contribution from the corona is smaller ($\lesssim2$\%). 

\citet{2017ApJ...841L...5S} interpreted a plasmoid ejection from a small flare kernel observed with AIA and ALMA Band\,3 (analogous to our first event) as emission from multi-thermal plasma where ALMA could trace a cooler ($10^4$\,K) component surrounded by an MK-hot sheet, but they noted that optically thin emission could be an alternative explanation. \citet{2019ApJ...875..163R} analyzed the same ALMA dataset and ruled out the latter, but suggested that the optical depth of the plasmoid is in the transition from optically thin to optically thick regime. 
Our analysis of the first event shows that the relative contribution from coronal plasma in the plasmoid is only approximately 2\% and is thus consistent with those previous observations. 

We also find that the core of the active region consists of long, warm mm fibrils that connect regions in the photosphere with strong opposite polarity field. This is in good agreement with synthetic mm maps that were obtained from a 3D r-MHD \texttt{Bifrost} simulation of an enhanced network \citep{2015A&A...575A..15L}. Even though that simulation is not meant to reproduce a solar active region, it shows the same kind of loop-like structures with $T_{\rm b}\lesssim$10\,kK at 3\,mm, albeit shorter than the observed ones, given the limited size of the computational box. They sampled relatively higher layers between $\sim$1500-2000\,km compared to the weakly magnetized areas. That simulation was also compared to a small-scale loop system observed with ALMA Band\,3 \citep{2020A&A...635A..71W}. 

The 3D r-MHD simulation that we used in this paper does not show the same kind of long mm fibrils (see Fig.\,\ref{fig:simulationOld}), likely because it lacks large-scale magnetic connectivity.
However, the process of magnetic flux emergence gives rise to localized heating events that resemble UVBs and FAFs. The brightness temperatures of the synthetic UVBs are in good agreement with the observed brightenings, but the simulation also produces a flaring fibril that is much brighter than any of the events that we observed. The 3\,mm continuum is optically thick at the base of the reconnection sites and traces the warm material in the upflowing high-velocity jets.

We note that our simulation does not include the effects of ambipolar diffusion, which has been shown to play an important role in 2.5D flux emergence experiments \citep[e.g.,][]{2006A&A...450..805L,2020ApJ...889...95M,2020A&A...633A..66N}, and it could affect the visibility of different heating events through changes in temperature and density. 
However, the 2.5D simulation of \citet{2020ApJ...891L...8M} that included the effects of ambipolar diffusion and nonequilibrium ionization of helium also suggests that ALMA Band\,3 will observe canopy fibrils that originate from strong concentrations of magnetic field, but it predicts low brightness temperatures ($\sim$4500--5000\,K) as consequence of expansion of cool dense plasma to much higher heights than in the 3D case \citep{2015A&A...575A..15L} constituting a cool canopy.  
The above is in contrast with our observations that show that the 3\,mm canopy within the active region is consistently warmer ($T_{\rm b}\sim$8000--9000\,K) than the QS ($T_{\rm b}\sim$7300\,K) and occasionally shows signs of impulsive heating and rapid flows ($\sim$40--340\,$\rm km\,s^{-1}$) of $T_{\rm b}\gtrsim10$\,kK plasma (or heat fronts) along fibrils (see Fig.\,\ref{fig:fafevol}) powered by NF kernels. 
Perhaps they are the signatures of precursor heating events of H$\alpha$ contrail fibrils \citep{2017A&A...597A.138R,2017A&A...598A..89R} caused by dissipation of electrical currents as in our \texttt{Bifrost} simulation.

\section{Conclusions}
\label{section:conclusions}

This work reports on the first results of a dedicated ALMA campaign to study the signature of small-scale heating events in the solar atmosphere using observations in the millimeter wavelength range. We compared SDO and ALMA Band\,3 (3\,mm) observations of an active region close to the disk center and we used a snapshot of a \texttt{Bifrost} 3D r-MHD simulation of flux emergence in order to investigate the visibility of reconnection events in the 3\,mm continuum.

We find that not even the brightest EB candidates identified in the AIA 1700\,\AA~continuum have a clear counterpart in Band\,3 that was observed at the highest spatial resolution so far (1.2\arcsec). 
This finding is consistent with the large body of evidence that EBs are photospheric reconnection phenomena \citep[e.g.,][]{Georgoulis_2002,2011ApJ...736...71W,Rutten_2013,2017A&A...598A..89R,2013ApJ...774...32V} and do not contribute significantly to heating the upper chromosphere. 

However, throughout the course of approximately one hour, we find multiple compact, bright, flickering mm-bursts (by analogy with the UV-bursts), and long fibrils that compose a warm canopy in the flux emergence region -- a distinct picture from the low-amplitude $T_{\rm b}$ variations in the QS.
The brightness temperatures of the kernels are typically 10\,kK, but reach as high as 14\,kK in the strongest events, and they last from dozens of seconds to several minutes. The thermal energy estimates place them in the category of nanoflare. Some of them could be UVBs, but IRIS observations would be required to confirm this. We also report for the first time on the detection of FAF-like events with Band\,3. We observe plane-of-sky motions of $T_{\rm b}$$\sim$9--12\,kK plasma with high horizontal velocities ($\sim$37--340\,$\rm km\,s^{-1}$) producing warm, fibril-like structures in the 3\,mm maps. One of the events shows the release of a blob-like structure from the flaring kernel similar to the plasmoid event as described in \citet{2017ApJ...841L...5S}.

The emissions in the EUV and mm are well correlated both in space and time, but we find a weak correlation between the bright UV patches and the 3\,mm structures. In particular, the lack of significant emission peak in the 1600\,\AA\ and 1700\,\AA~filters in most cases suggests that the mm-bursts occur in higher layers of the atmosphere. 

Given the increase of the emission measure at the NF kernels a significant contribution of several hundred kelvin ($\approx$4-7\%) from the lower corona may overlap with the chromospheric signal the Band\,3 as shown by the DEM analysis. 
The remarkable spatial and temporal correlations between the 3\,mm continuum and AIA 304\,\AA~also suggest that Band\,3 may be sensitive to transition region temperatures at those locations.

Our simulation shows that, in general, ALMA may be able to detect the mm analogues of UVBs and small flares should they cause a significant temperature increase in the chromosphere. Therefore, it is possible to use mm continuum observations to directly estimate the temperature of the reconnection and constrain their formation heights. We note that our analysis does not allow us to conclude that magnetic reconnection is the cause of the observed brightenings.

Future work should also aim to further investigate the contribution functions of the mm continuum under flaring conditions that are very different from the QS, where the 3\,mm emission is predominantly chromospheric.
Our work shows that ALMA observations can be used to detect localized, transient heating phenomena in active regions and possibly constrain magnetic reconnection modeling, while comparisons with other diagnostics provide interesting insight into the thermal structure of the solar atmosphere.

\begin{acknowledgements}

We thank C. Froment for the comments on the paper. This paper makes use of the following ALMA data: ADS/JAO.ALMA\#2018.1.01518.S, ALMA is a partnership of ESO (representing its member states), NSF (USA) and NINS (Japan), together with NRC (Canada), MOST and ASIAA (Taiwan), and KASI (Republic of Korea), in cooperation with the Republic of Chile. The Joint ALMA Observatory is operated by ESO, AUI/NRAO and NAOJ.

The Institute for Solar Physics is supported by a grant for research infrastructures of national importance from the Swedish Research Council (registration number 2017-00625).

The forward calculations were performed on resources provided by the Swedish National Infrastructure for Computing (SNIC) the National Supercomputer Centre (NSC) at Link\"{o}ping University.

JdlCR is supported by grants from the Swedish Research Council (2015-03994), the Swedish National Space Agency (128/15) and the Swedish Civil Contingencies Agency (MSB). This project has received funding from the European Research Council (ERC) under the European Union's Horizon 2020 research and innovation programme (SUNMAG, grant agreement 759548).

SW was supported in this work by a grant from AFOSR.

JL is supported through the CHROMATIC project (2016.0019) funded by the Knut and Alice Wallenberg foundation.

GV is supported by a grant from the Swedish Civil Contingencies Agency (MSB).

This research was supported by the Research Council of Norway through grant 170935/V30, through its Centres of Excellence scheme, project number 262622. Computing time has come through grants from the Norwegian Programme for Supercomputing,  as well as from the Pleiades cluster through the computing project s1061, from the High End Computing (HEC) division of NASA.

This research has made use of \texttt{Astropy}\footnote{\url{https://astropy.org}} -- a community-developed core Python package for Astronomy \citepads{astropy:2018}, \texttt{SunPy}\footnote{\url{https://sunpy.org}} -- an open-source and free community-developed solar data analysis Python package \citep{2015CS&D....8a4009S}, and MATLAB and the Image Processing Toolbox release 2018b (The Mathworks, Inc., Natick, MA, USA).

\end{acknowledgements}

\begin{appendix}

\section{Millimeter brightness oscillations}
\label{section:oscillations}

\begin{figure}
    \centering
    \includegraphics[width=\linewidth]{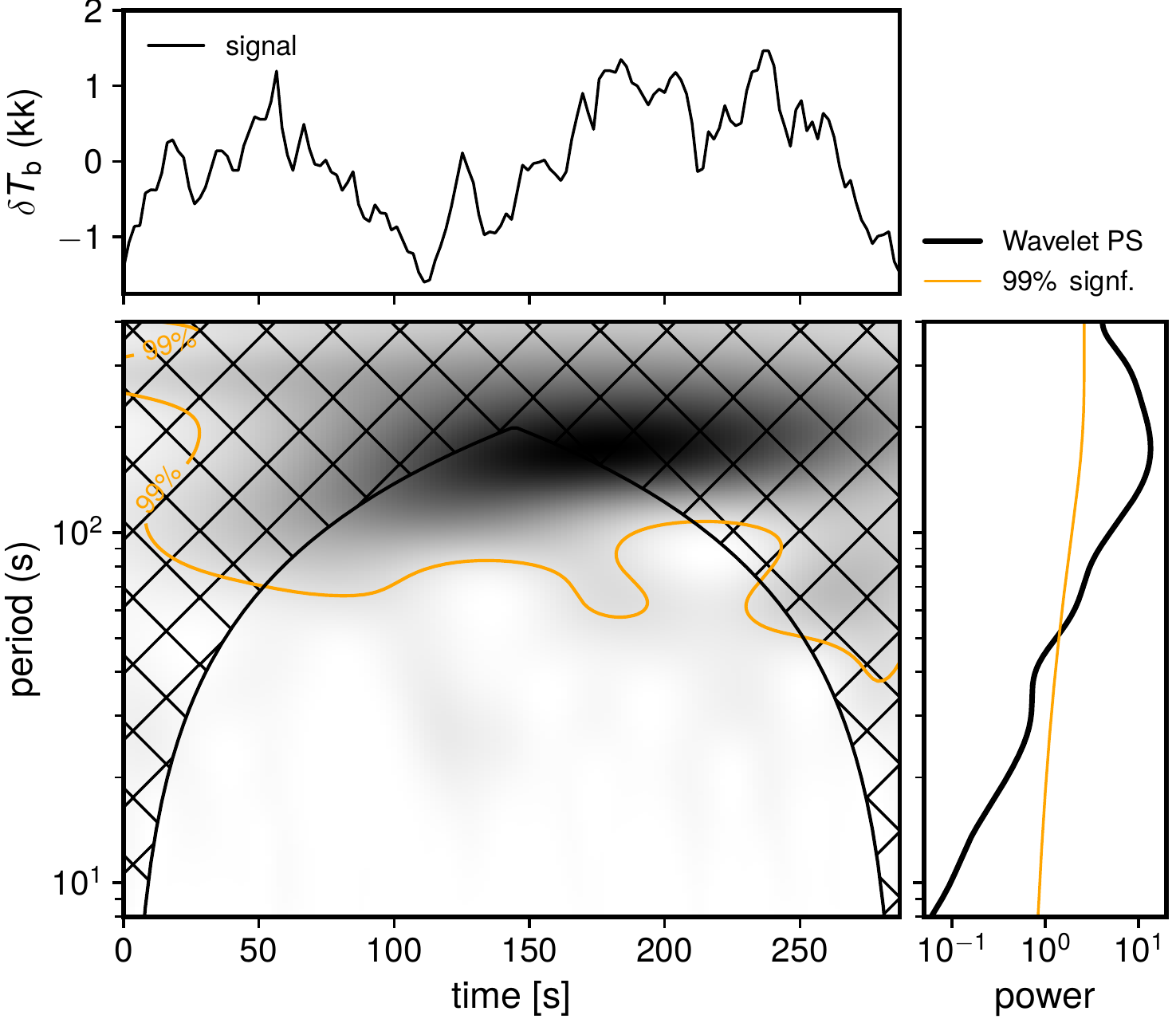}
    \caption{Time-frequency domain analysis for NF1. The scalogram is shown in inverse color. The hatch is the cone of influence. The 99\% confidence regions are shown by the yellow lines. The global wavelet power spectrum (PS) and global significant level are shown in the right panel.} 
    \label{fig:wavelet1}
\end{figure}

\begin{figure}
    \centering
    \includegraphics[width=\linewidth]{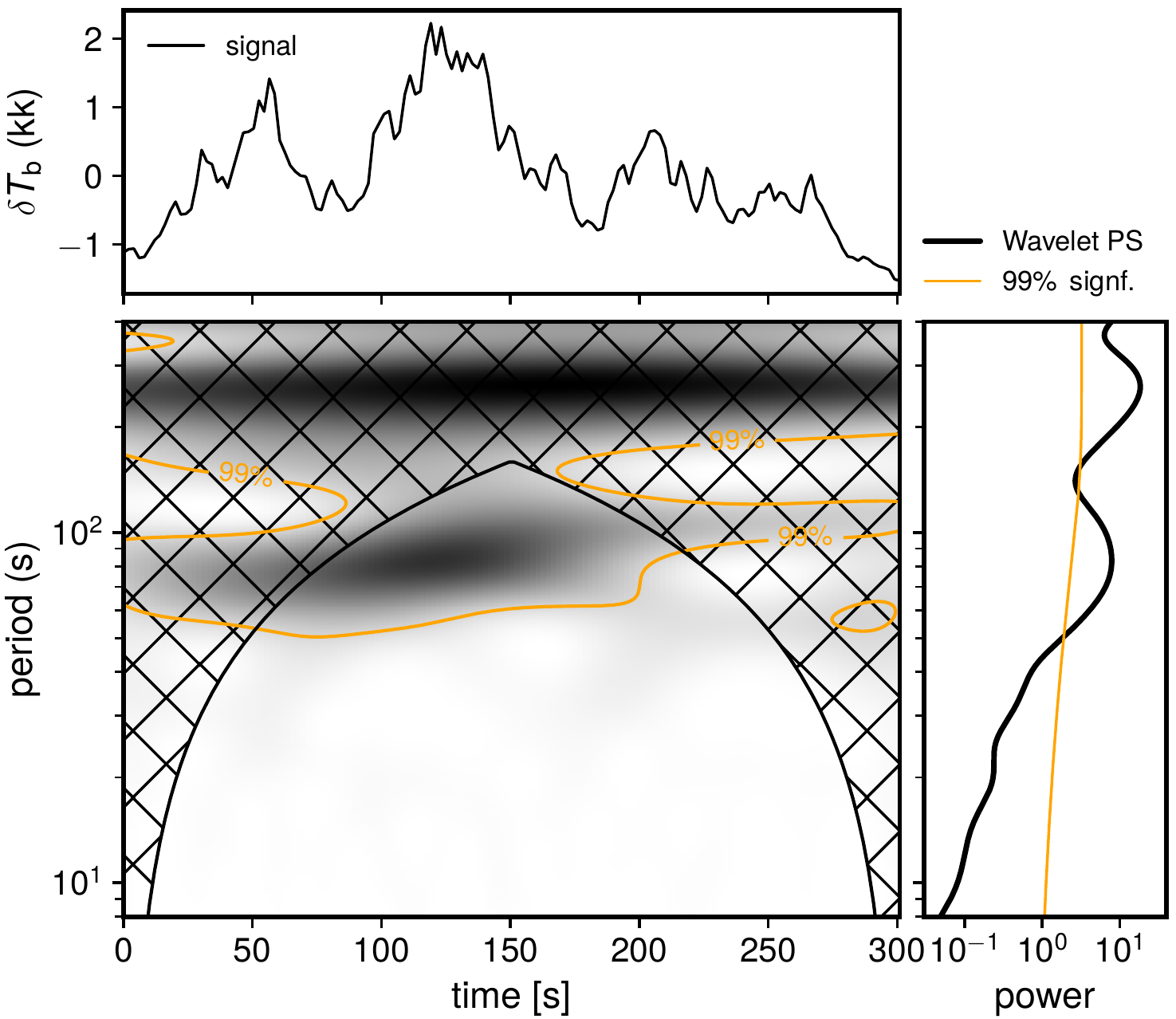}
    \caption{Time-frequency domain analysis for NF3. Analogous to Fig.\,\ref{fig:wavelet1}.}
    \label{fig:wavelet2}
\end{figure}

High-amplitude $T_{\rm b}$ variations of several hundred kelvin stand out in the high-cadence ALMA lightcurves in Fig.\,\ref{fig:lcurves}. 
In order to investigate whether there are any periodicities in the brightenings in the two main events (see Fig.\,\ref{fig:fafevol}) we conducted time-frequency analysis using the Python wavelet software\footnote{\url{https://github.com/chris-torrence/wavelets}} based on \citet{Torrence98}. We focus on the two signals with longest duration (NF1 and NF3) at the center of the brightenings. No detrending is performed in order to avoid spurious frequencies in the power spectra \citep[e.g.,][]{2016ApJ...825..110A}.

The results of the wavelet analysis are shown in Fig. \ref{fig:wavelet1} and Fig.\,\ref{fig:wavelet2} for NF1 and NF3, respectively. The cone of influence represents the region where boundary effects may affect the wavelet coefficients.
We find evidence for periodicities in the $\approx$50-150\,s range in NF1 and NF3 at a 99\% confidence level assuming white noise. We find a prominent power peak at $\approx$80\,s in NF3.

We investigated whether the Band\,3 jitter (see Sect.\,\ref{section:observations}) has an effect on the observed oscillations by averaging the ALMA sequence in a 5$\times$5 px box centered on the NFs and found that the shape of the signals is practically unaltered besides the obvious lowering of their average value. This has no significant effect on their power spectra. In this regard, we note that the 3\,mm signals correlate with the EUV intensity variations (see Fig.\,\ref{fig:lcurves}).

\citet{2015ApJ...809...82G} reported on similar frequencies derived from IRIS observations of UVBs and links them to changes in the emerging photospheric magnetic field. We do not find evidence for the latter. Intensity fluctuations in visible and UV diagnostics have also been associated to plasmoid-mediated reconnection \citep[e.g.,][]{2015ApJ...813...86I,2017ApJ...851L...6R,2019A&A...628A...8P}. Apart from the intermittency of the mm signals, the release of a fast-moving blob from NF1 in the first event is tentative evidence for such scenario. 
We speculate that the lack of a similar feature in NF3 could be due to insufficient spatial resolution since plasmoids have been found at scales ten times smaller \citep{2017ApJ...851L...6R} than what we can resolve with ALMA Band\,3. Therefore, the signals may be caused by a sequence of reconnection events. This warrants further investigation.

\section{Supplementary figures}
\label{section:extraFigs}

\begin{figure*}
    \centering
    \sidecaption
    \includegraphics[width=120mm]{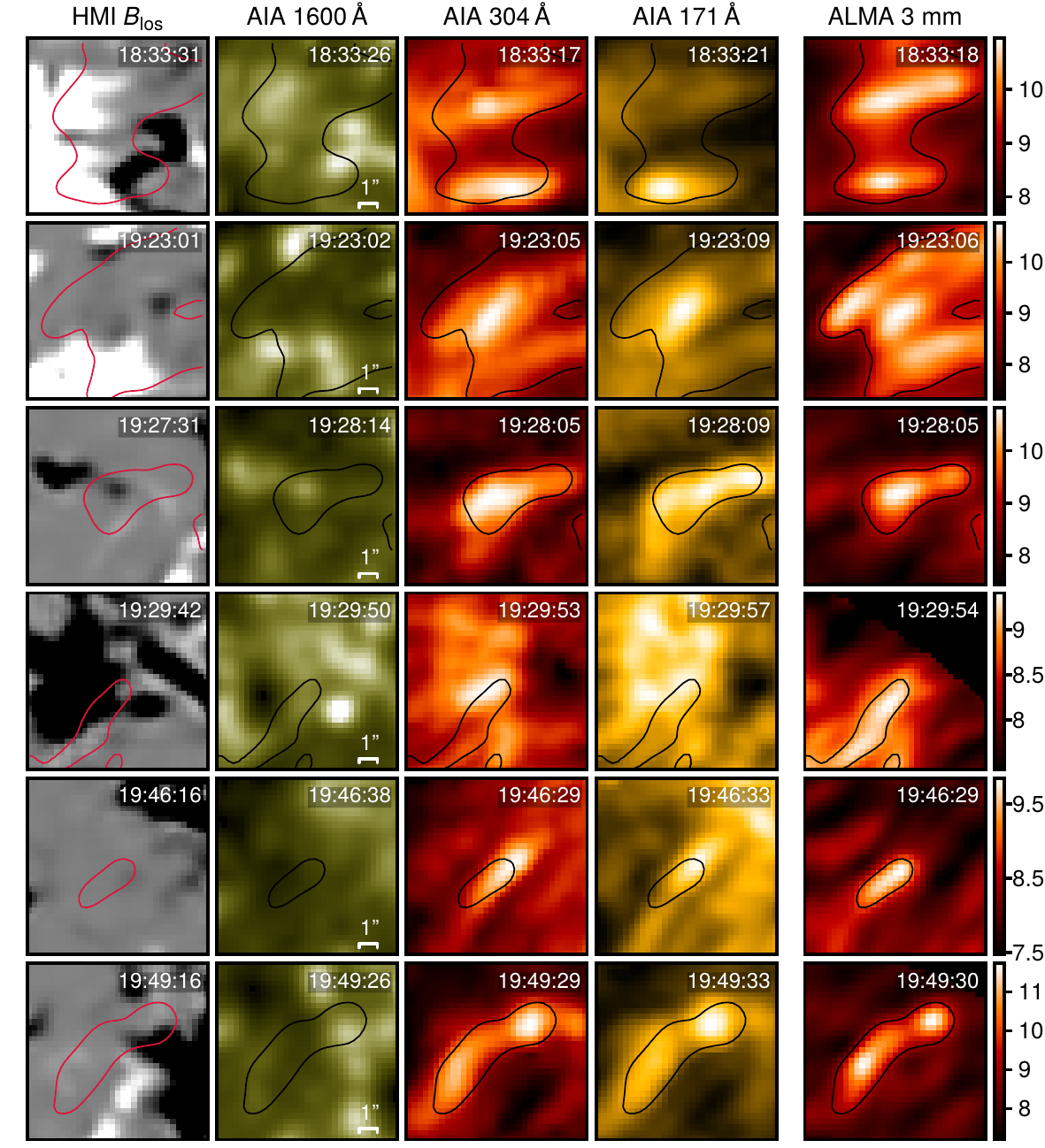}
    \caption{ Other ALMA Band\,3 counterparts of EUV brightenings. From left to right: HMI magnetogram, spectral radiance in AIA\,1600\,\AA, AIA\,304\,\AA, and AIA\,171\,\AA, and ALMA 3\,mm brightness temperature. All panels show a 10\arcsec$\times$10\arcsec~FOV centered on the bright structures. The contours correspond to $T_{\rm b}(3\,\mathrm{mm})=9$\,kK. The image scale is indicated in the panels in the second column. 
    The range in the HMI magnetograms is clipped at $\pm0.3$\,kG. The EUV images are displayed in logarithmic scale and the colormaps are individually scaled.
    The ALMA color bars are in units of kilokelvin.}
    \label{fig:mfpanelextra}
\end{figure*}

This appendix contains supplementary figures with SDO and ALMA images of the remaining EB candidates (Fig.\,\ref{fig:ebpanelextra}) and other EUV brightenings (Fig.\,\ref{fig:mfpanelextra}) with a layout similar to Fig.\,\ref{fig:ebpanel} and Fig.\,\ref{fig:mfpanel}, respectively. As in the main events described in Sects.\,\ref{section:EBsandMFs} and \ref{section:DEM}, no Band\,3 counterpart of strong 1700\,\AA~EB emission is detected, but we found different compact or jet-like mm-bursts with cospatial EUV emission.

\begin{figure*}
    \centering
    \sidecaption
    \includegraphics[width=120mm]{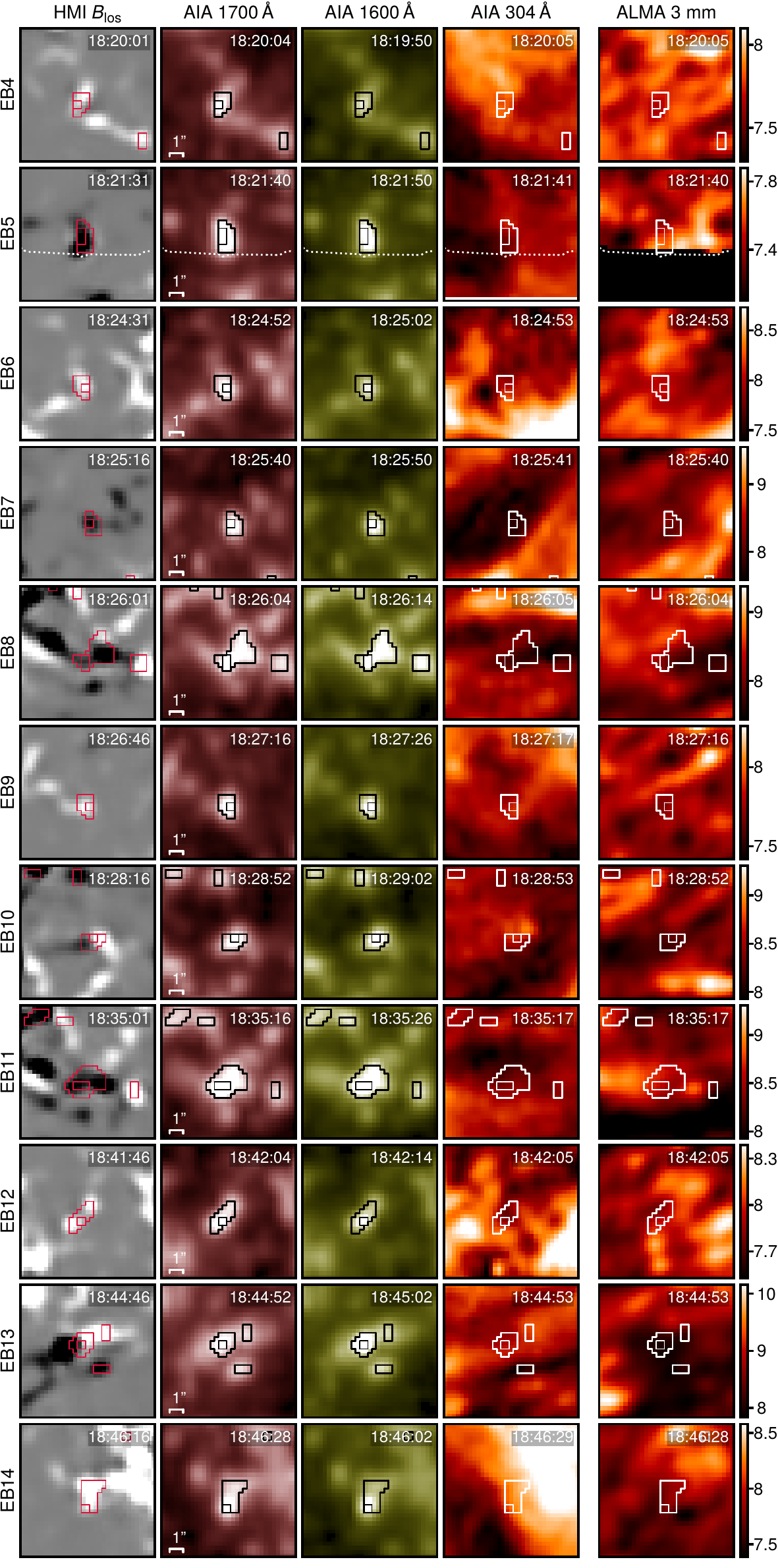} 
    \caption{EB candidates in SDO and ALMA. All panels show a 10\arcsec$\times$10\arcsec~FOV centered on EBs. The contours correspond to $5\sigma$ (thick) and $9\sigma$ (thin) \texttt{EBDETECT} thresholds (see Sect.\,\ref{section:EBsandMFs}). Only the $9\sigma$ events are considered as EBs. The image scale is indicated in the panels in the second column. The dotted lines in the second row delimit the edge of the ALMA field of view.
    The range in the HMI magnetograms is clipped at $\pm0.3$\,kG and the intensities in the AIA\,1700\,\AA, 1600\,\AA~and 304\,\AA~channels are capped at 4000\,$\rm DN\,s^{-1}$, 150\,$\rm DN\,s^{-1}$, and 3000\,$\rm DN\,s^{-1}$ in all panels. The ALMA color bars are in units of kilokelvin.}
    \label{fig:ebpanelextra}
\end{figure*}
\begin{figure*}
    \ContinuedFloat
    \centering
    \sidecaption
    \includegraphics[width=120mm]{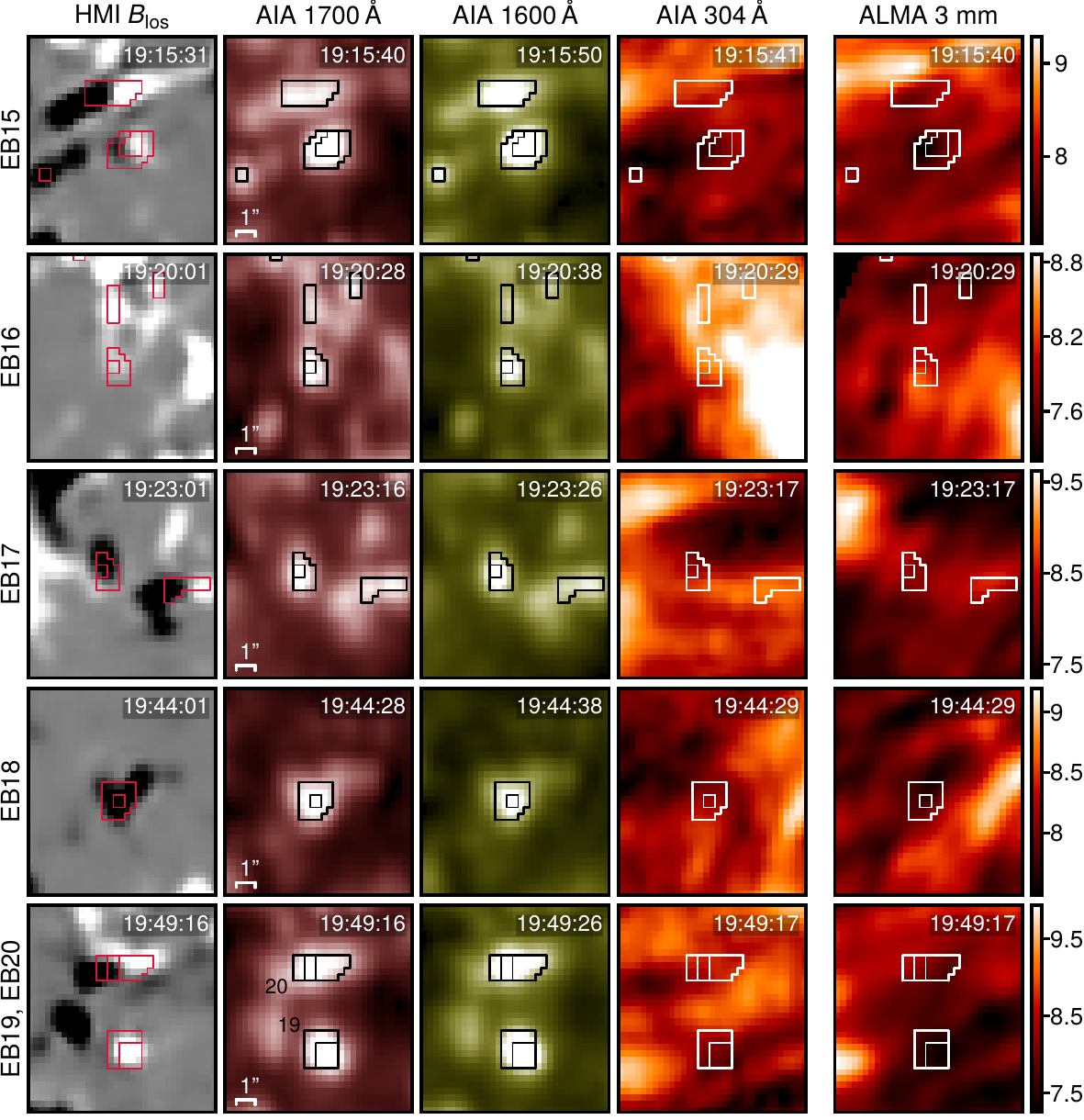} 
    \caption{Continued.} \label{}
\end{figure*}

\end{appendix}

%
\bibliographystyle{aa} 

%

\end{document}